\documentclass{aa}  

\usepackage{graphicx}
\usepackage{txfonts}
\usepackage{subfig}
\usepackage{threeparttable}
\usepackage{longtable,lscape}
\usepackage{natbib}
\usepackage{fixltx2e}

\begin{document}

   \title{Archival VLT/NaCo multiplicity investigation of exoplanet host stars}

   \author{J. Dietrich
          \inst{1,2}
          \and
          C. Ginski\inst{1,3}
          }

   \institute{Leiden Observatory, Leiden University, P.O. Box 9513, 2300 RA Leiden, The Netherlands\\
              \email{jeremy.dietrich@cfa.harvard.edu}
         \and
             Harvard University, Cambridge, MA 02138, United States
         \and Anton Pannekoek Institute for Astronomy, University of Amsterdam, Science Park 904, 1098 XH Amsterdam, The Netherlands
             }

   \date{Received June XX, 2017; accepted YYY}

 
  \abstract
   {The influence of stellar multiplicity on planet formation is not yet well determined. Most planets are found using indirect detection methods via the small radial velocity or photometric variations of the primary star. These indirect detection methods are not sensitive to wide stellar companions.
   High-resolution imaging is thus needed to identify potential (sub)stellar companions to these stars.}
   {In this study we aim to determine the (sub)stellar multiplicity status of exoplanet host stars, that were not previously investigated for stellar multiplicity in the literature.
   For systems with non-detections we provide detailed detection limits to make them accessible for further statistical analysis.}
   {For this purpose we have employed previously unpublished high-resolution imaging data taken with VLT/NACO in a wide variety of different scientific programs and publicly accessible in the ESO archive.
   We used astrometric and theoretical population synthesis to determine whether detected companion candidates are likely to be bound or are merely chance-projected background objects.}
   {We provide detailed detection limits for 39 systems and investigate 29 previously unknown companion candidates around five systems. In addition, we show for the first time that the previously known companion candidate around HD\,204313 is likely a background object. 
By comparison with secondary epochs of 2MASS data we show that the companion candidates around GJ176 and HD\,40307, as well as two of the sources around HD\,85390, are likely background objects. For HD\,113538 and HD\,190984, as well as multiple further companion candidates around HD\,85390, further observational data is required to test common proper motion of the companion candidates.}
   {}

   \keywords{   techniques: high angular resolution --
                binaries: visual --
                planets and satellites: formation
               }

   \maketitle
%

\section{Introduction}

The past decade has seen a rapid growth of the number of known exoplanet systems. Today more than 1200 such systems are confirmed (exoplanet.eu, \citealt{2011A&A...532A..79S}). The overwhelming majority of these systems were discovered using two indirect detection methods: the radial velocity method and the transit method. The radial velocity method finds planets by detecting periodic Doppler shifts in their host stars' spectral lines, while the transit method searches for the periodic dimming of the host star due to the planet transiting in front of it. While both methods are sensitive to exoplanets within fractions of and up to a few au, they are mostly blind for stellar companions at wider separations. A large proportion of the stars in the Galaxy are part of multiple stellar systems. \cite{2010ApJS..190....1R} found that 46$\pm$2\% of all solar type stars within 25\,pc are members of stellar multiple systems. In fact, \cite{2012A&A...542A..92R} point out that the first evidence for an extrasolar planet was indeed found around the young binary star $\gamma$\,Cep by \cite{1988ApJ...331..902C}. This planet was later confirmed by \cite{2003ApJ...599.1383H}. It is thus desirable to understand the implications of stellar multiplicity on the planet formation process. Theoretical studies show that a stellar companion might truncate or stir the planet-forming disk (see e.g., \citealt{1994ApJ...421..651A}, \citealt{2005MNRAS.363..641M}). For an extensive overview of the effects of stellar companions on disks, see \cite{2014arXiv1406.1357T}. In addition, stellar companions might influence the dynamics of formed planets, for example, by altering inclination and eccentricity of their orbits via mutual Kozai-Lidov type resonances (see e.g., \citealt{2003ApJ...589..605W}). \\
   To understand the dominant effects that stellar companions exercise on planetary systems it is necessary to study their stellar multiplicity. This can be done by high-resolution seeing or diffraction limited imaging. The number of such studies has been increasing in the past few years, such as \cite{2006A&A...456.1165C}, \cite{2007A&A...474..273E}, \cite{2012MNRAS.421.2498G}, or more recently \cite{2014A&A...566A.103L}, \cite{2015MNRAS.450.3127M}, \cite{2015A&A...579A.129W}, \cite{2016MNRAS.457.2173G}, and \cite{2017AJ....153..242N}. These studies were carried out with telescopes in the 3\,m to 8\,m class - telescopes that are usually in high demand for a variety of science programs. It is thus necessary to make such surveys as efficient as possible. In particular, it is important that all available high-resolution imaging data of exoplanet host stars is analyzed and that detection limits are presented even in cases where no detection was made. This prevents inefficient re-observation of the same targets, and provides valuable statistical constraints on multiplicity of exoplanet hosts. The ESO archive offers a large amount of such data, which were taken with various scientific goals, sometimes even before the observed star was known to host an exoplanet. \\
   In this work we present our results of archival VLT/NaCo (\citealt{2003SPIE.4841..944L}, \citealt{2003SPIE.4839..140R}) data. In the following sections we state our data selection criteria, describe the available data and the subsequent data reduction, and present measurements for companion candidates and detection limits for each data set.

   \section{Archival data description and reduction}

We used a list of all confirmed exoplanet hosts detected by radial velocity and/or transit method compiled from exoplanet.eu (\citealt{2011A&A...532A..79S}) as input for our search. In order to narrow down our sample size with restrictions based on observing with VLT/NaCo, we excluded all stars north of +30 degrees in declination. Then, we searched the VLT/NaCo archive for each system and excluded all systems that did not appear in the archive. For all systems with available archival data we checked for previous publications of this data that were: linked to the program in the ESO archive, listed at exoplanet.eu, listed in the SIMBAD database, or given in ADS for the specific target using the host star
name given in the ESO archive as input. 

The title and/or keywords of all recovered publications were then scanned for keywords such as ``direct imaging'', ``lucky imaging'', ``high-contrast imaging'' or ``multiplicity study''. If the retrieved VLT/NaCo archival data for each system was discovered to already be in publication, the system was subsequently removed from our target list. The remaining systems became our sample, consisting of 46 sets of observations comprising 39 stars with exoplanets.The observations ranged in wavelength from the H band to the L\_prime band, as well as many narrow bands. Some of the images also used neutral density overlays. Six of the systems observed in the H band were done in spectral differential imaging (SDI) mode with a Wollaston Prism splitting the light into multiple narrow bands in the same wavelength range. In addition, two of the systems were imaged using angular differential imaging (ADI). We also searched for observations for these systems in other AO archives (Gemini, Keck), but there was no data available for this specific set. A summary of all observations is shown in Table~\ref{tab:VLT_archives_table}.

We developed a Python pipeline utilizing PyFITS (\citealt{1999ASPC..172..483B}) to perform our own reduction and analysis of the images, including possible candidate detection limits. The pipeline takes in a set of images, both science and calibration images, and performs a reduction and combination of the images. During this process, it performs a median frequency filter to lessen striping effects. It then creates a normalized master flat field from the flat calibration images by combining them, averaging pixel-by-pixel, and dividing by the median value. After flat-fielding all the science frames, the pipeline creates a median sky field by combining each flat-fielded science image and taking the median pixel-by-pixel. Once it subtracts the median sky image from the science frames, the program shifts and adds the sky-subtracted images using cumulative offset header data to create one master reduced science image.\\
  For the six systems imaged with the SDI Wollason prism, we created a Python routine that performed the same flat-fielding as before, but it then split the four quadrants of the image into separate pieces. This Python routine then subtracted the median sky field and shifted them so the target star was in the center of the new, smaller image, before adding them together as one image. However, this routine does not use the full inherent capabilities of SDI Wollaston imaging, such as the on-feature and off-feature narrow bands used to detect emission features like methane found in relatively high abundances in exoplanetary atmospheres but are almost nonexistent in stellar atmospheres.\\
  Each of the remaining two systems imaged in ADI mode required a separate Python program. HD 98649 was also jittered, so after flat-fielding our routine shifted all the jittered images so the target star was centered in the image, and then derotated them by the parallactic angle. Finally, it subtracted the median sky value and added the images together. HD 40307 was imaged in ADI mode with multiple rotational offsets rather than jitters. Our Python program had to derotate by the telescope pointing angle first, before subtracting the median sky and finally derotating by the parallactic angle.

\section{Detected companion candidates}  

From our 39 sample systems, we identified companion candidates (CCs) for six systems: GJ 176, HD 40307, HD 85390 (12 candidates), HD 113538 (12 candidates), HD 190984, and HD 204313. We show the reduced images of these systems in Fig.~\ref{fig:companion_candidates_VLT_figure}. In addition, we show close-up signal-to-noise ratio (S/N)\ maps of each companion candidate in the appendix.
We searched through other observational archives that include high contrast imaging as well as survey data, including Keck, Gemini, Subaru, and 2MASS, but we only found suitable images for five candidates from another observing epoch in VLT/NaCo or 2MASS on which astrometry and photometry could be done: GJ 176, HD 40307, HD 85390 (two possible candidates), and HD 204313. The relevant 2MASS images are shown in Fig~\ref{fig:companion_candidates_2MASS_figure}. \\

\subsection{Photometric measurements}

For both the VLT/NaCo images and the 2MASS images we used the aperture photometry tool (APT, \citealt{2012PASP..124..737L}) to measure the absolute difference in magnitude between the main target star and the companion candidate. In APT, we adjusted the inner aperture ring of the selection to include all of the main (bright) star, while also insuring that the outer ring was well into the background, and that the middle annulus contained all possible contamination sources. After APT calculated a magnitude for the main star, we then moved the aperture to the companion candidate and measured that magnitude, using the exact same selection. This gave us a magnitude for the companion candidate. We were interested in deriving the relative contrast between the candidates and the primary from the NaCo observations that could then be translated to absolute magnitudes from the 2MASS magnitudes of the primary star, the related spectral types, and the corresponding filter transmissions. These specific magnitudes were not calibrated to be exact, but the absolute difference in magnitude was correct, which was our interest. 
Once we calculated the absolute difference in magnitude, we then applied the 2MASS (\citealt{2006AJ....131.1163S}) magnitude in the corresponding filter and calculated the true apparent magnitude for the candidate companion which we give in Table~\ref{tab:VLT_data_table}.\\

\subsection{Astrometric measurements and proper motion analysis}
Using ESO MIDAS (Munich Image Data Analysis System, \citealt{1983Msngr..31...26B}) and a custom Python routine, we performed astrometry with sub-pixel accuracy to find the exact centers of both the main star and each analyzable companion candidate to calculate the separation in pixels and the position angle between the two stars. To find the exact centers, when applicable, we used the ``center/Gauss'' function in ESO MIDAS; otherwise, for images with a low S/N detection where a Gaussian cannot be fitted, we used the ``center of mass'' method in NumPy to find a less-accurate value for the star's center. 
The separation was then converted to arcseconds using the pixel scale of the images, which was found in the headers. To account for the uncertainty of this astrometric calibration we used typical calibration uncertainties for VLT/NACO as reported in \cite{2010A&A...509A..52C} and \cite{2014MNRAS.444.2280G}.
We find an average error for true north of 0.17\,deg. For the pixel scale we find uncertainties of 0.04\,mas/pixel for the S13 and SDI+ scale as well as 0.05\,mas/pixel for the S27 scale.
Values and 1-$\sigma$ errors were calculated by center/Gauss and center of mass, and were propagated throughout the calculation. Results are shown in Table~\ref{tab:VLT_data_table} for VLT/NaCo and in Table~\ref{tab:2MASS_data_table} for 2MASS epochs.\\
  For the four companion candidates with two observing epochs we performed a common proper motion analysis. To determine the possible change in separation or position angle due to slow orbital motion, we used the distance measurements together with the separation in arcseconds to determine the physical separation in au if the companion was comoving, then calculated the period which the companion would rotate assuming that the physical separation measured is the semi-major axis of the orbit. We then extrapolated backward from the VLT/NaCo dataset to find an earlier position angle and separation of a comoving orbit, and then checked the values we obtained from the 2MASS dataset to determine whether the candidates are actual companions or just chance background stars. Further explanation of this common method can be found in, for example, \citet{2012MNRAS.421.2498G}.\\
   The results are shown in Figs.~\ref{fig:Analysis_figure_1}, \ref{fig:hd40}, \ref{fig:Analysis_figure_3}, and \ref{fig:Analysis_figure_4}. We expect near constant astrometric positions in both epochs for comoving objects (with potential slow orbital motion as described earlier), while background objects should show a behavior consistent with the "wobbled" lines. The periodic wobble is caused by the discussed parallactic displacement.\\

\subsubsection{GJ\,176}
\begin{figure*}
        \centering
        \subfloat[]{\includegraphics[width=.5\textwidth]{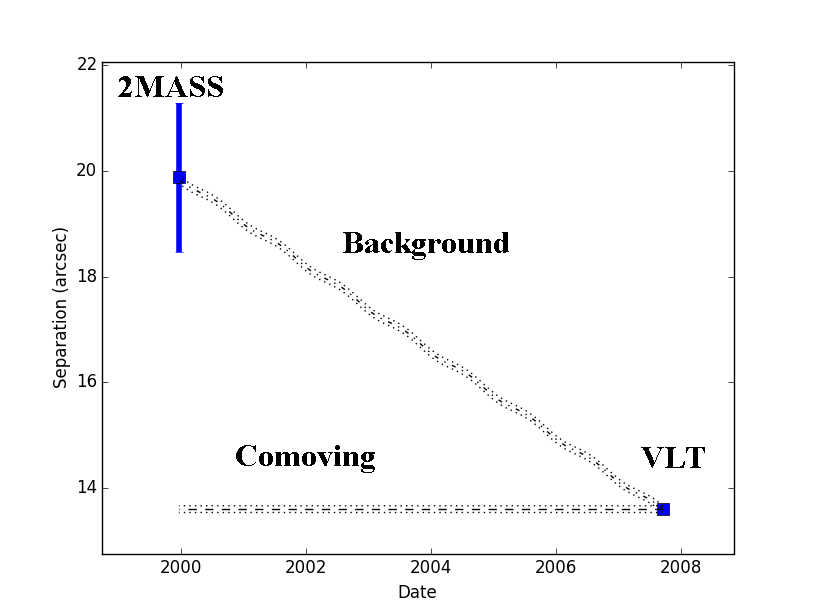}}
        \subfloat[]{\includegraphics[width=.5\textwidth]{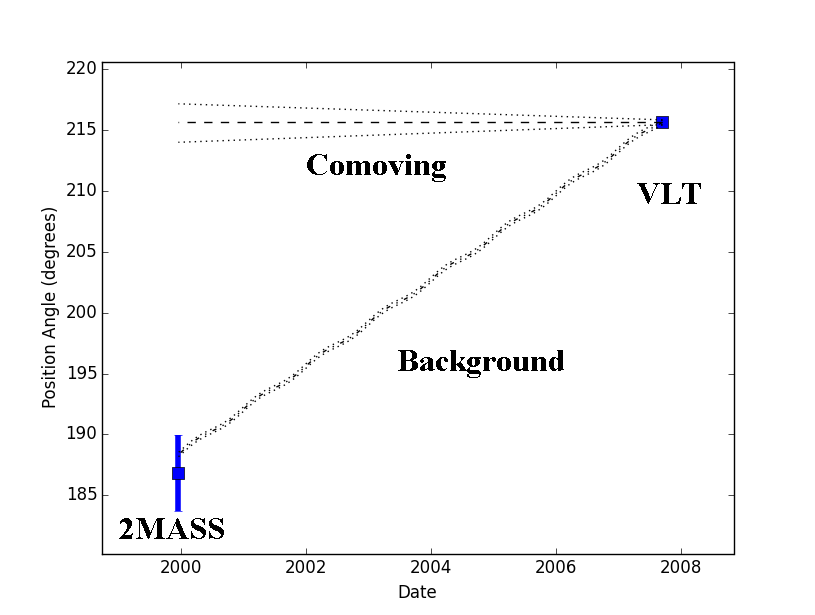}}
    \caption{Separation (left panel) and position angle (right panel) analysis for GJ 176 CC.}
    \label{fig:Analysis_figure_1}
\end{figure*}

For the GJ\,176 system (Fig.~\ref{fig:Analysis_figure_1}), the 2MASS and NACO astrometric epochs are perfectly consistent with the companion candidate being a nonmoving background object.

\subsubsection{HD\,40307}
\begin{figure*}
        \centering
        \subfloat[]{\includegraphics[width=.5\textwidth]{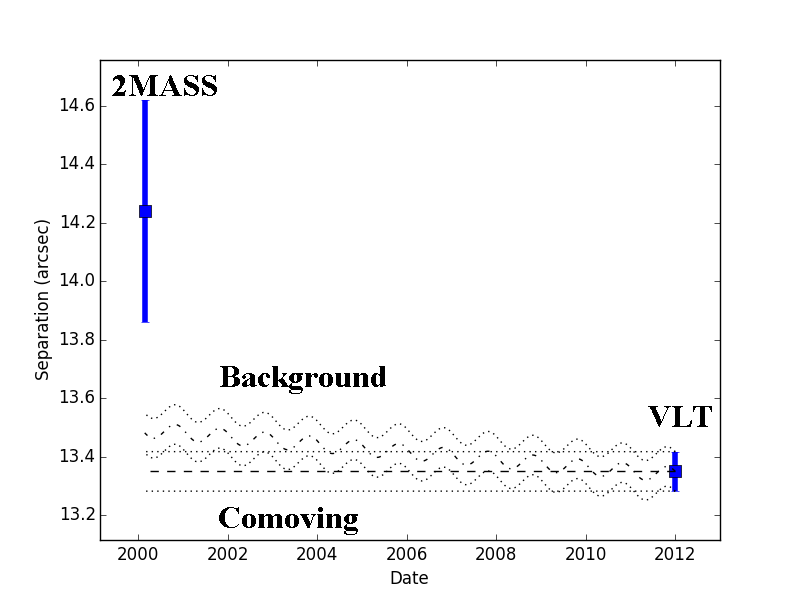}}
        \subfloat[]{\includegraphics[width=.5\textwidth]{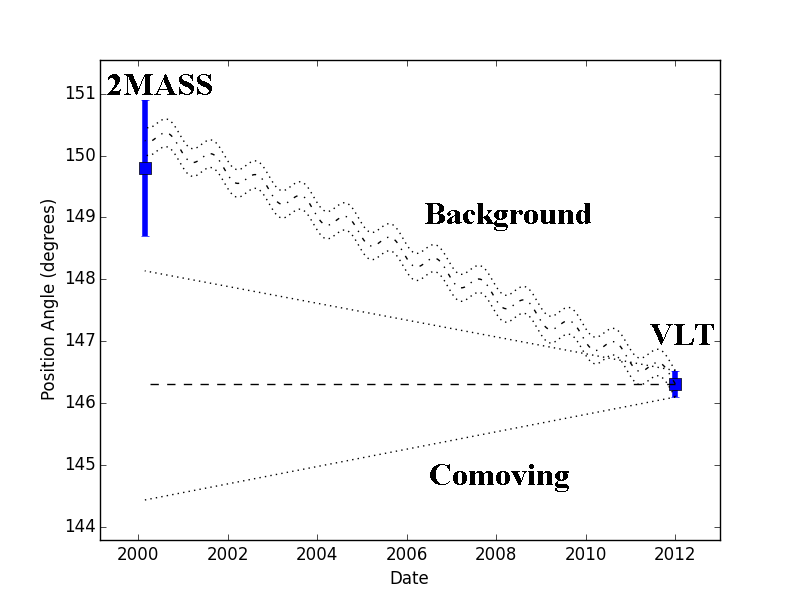}}
    \caption{Separation (left panel) and position angle (right panel) analysis for HD\,40307 CC.}
    \label{fig:hd40}
\end{figure*}

While the position angle extracted from the 2MASS image for the HD\,40307 system (Fig.~\ref{fig:hd40}) is consistent with a background object, the separation is in principle too large. This could be explained by the fact that in this case we used unsharp masking of the 2MASS image before extracting the astrometry, since the position of the faint companion candidate was strongly dominated by the bright halo of the primary star. For this purpose we convolved the original 2MASS image with a Gaussian kernel and then subtracted the resulting image from the original. While this allowed a clearer detection of the companion candidate it likely also oversubtracted flux in the direction of the primary star. This probably lead to a slight shift of the center of the faint source away from the primary star, and thus explains the slightly too large separation. In any case, neither the change of separation nor of position angle could be explained by slow orbital motion and is more consistent with a non-moving background object. We thus conclude that the companion candidate to HD\,40307 is indeed a chance aligned background object.\\

\subsubsection{HD\,85390}
\begin{figure*}
        \subfloat[GJ 176]{\includegraphics[width=.33\textwidth]{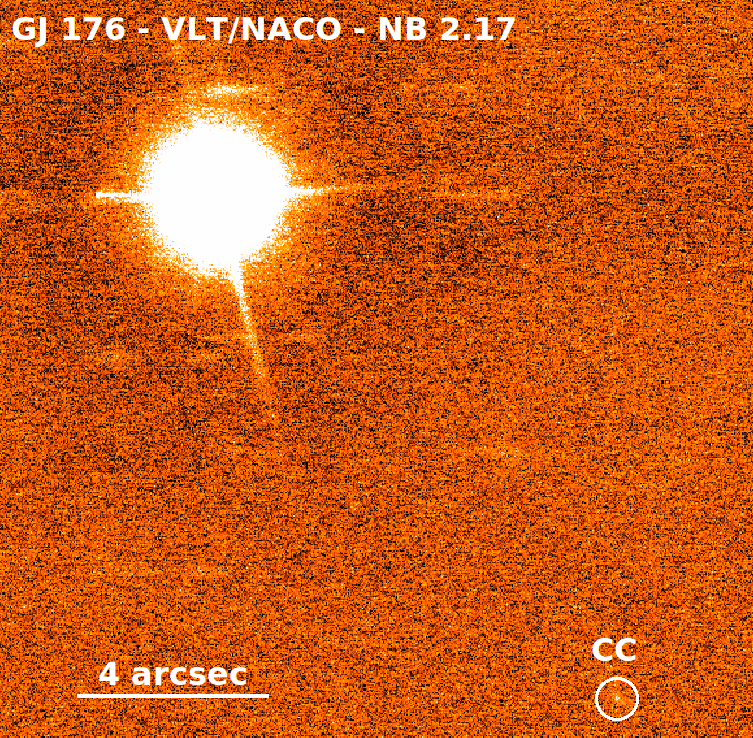}}
        \subfloat[HD 40307]{\includegraphics[width=.33\textwidth]{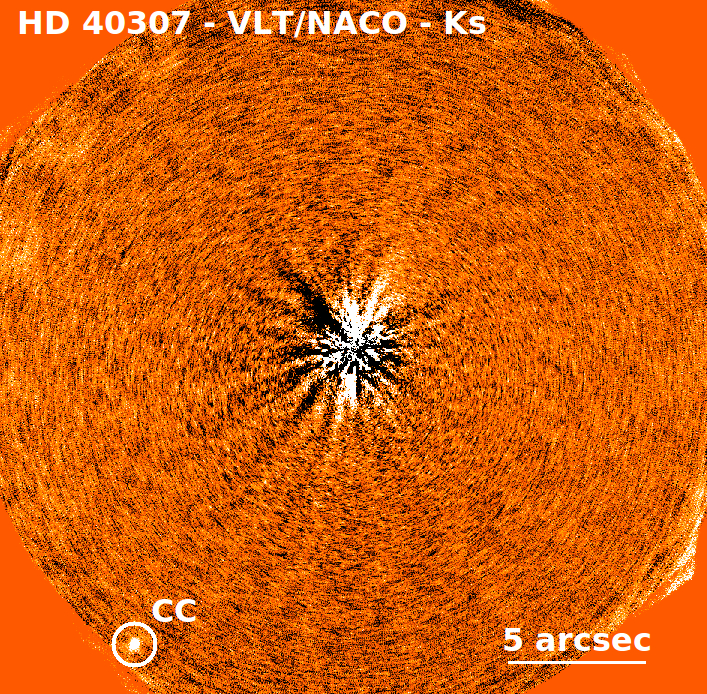}}
        \subfloat[HD 85390]{\includegraphics[width=.33\textwidth]{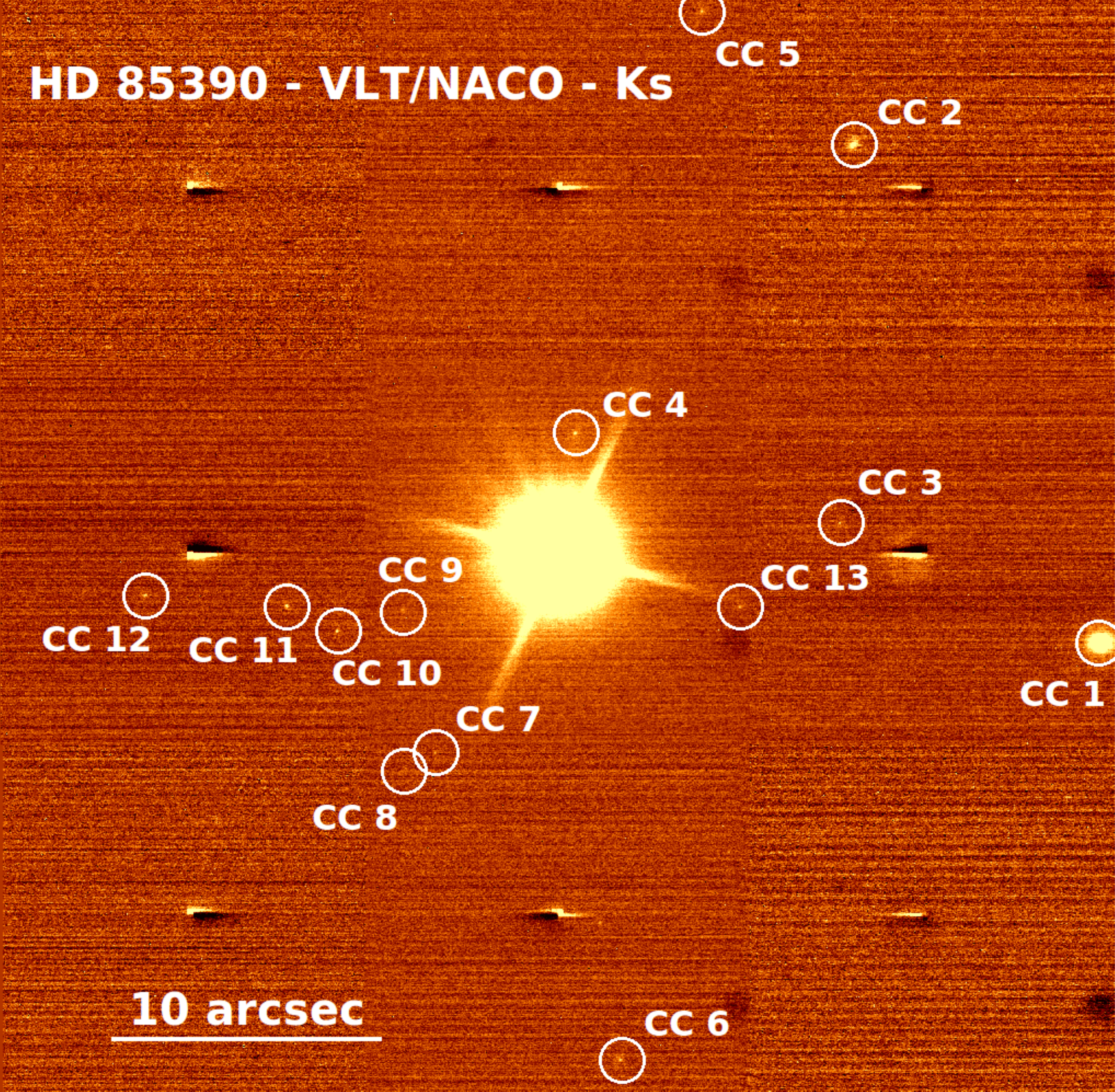}\label{hd85390_naco}}
        
        \subfloat[HD 113538]{\includegraphics[width=.33\textwidth]{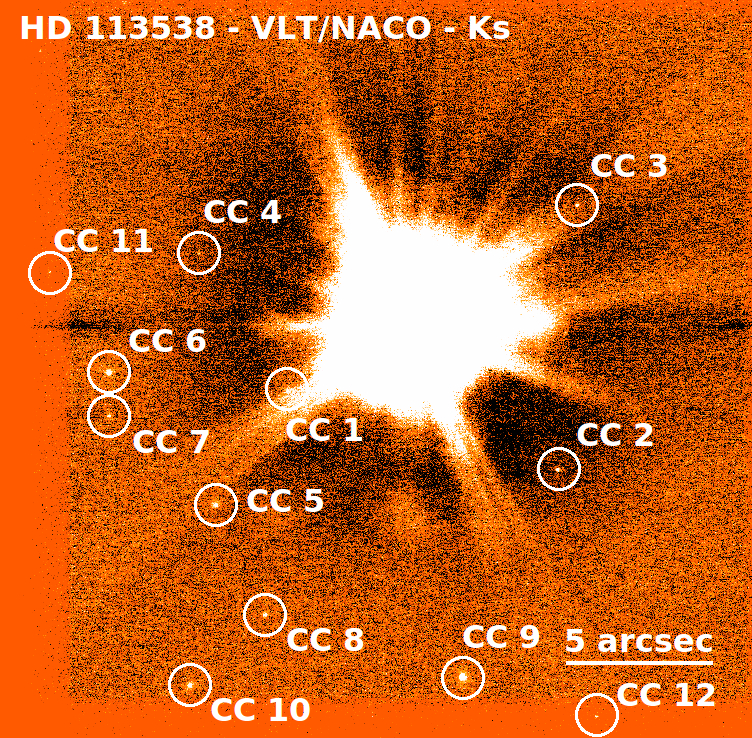}}
        \subfloat[HD 190984]{\includegraphics[width=.33\textwidth]{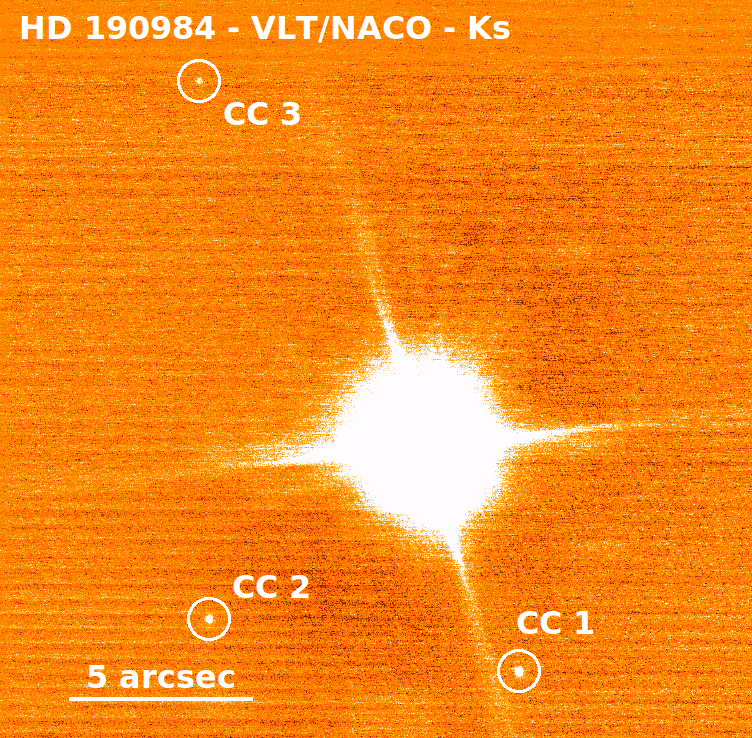}}
        \subfloat[HD 204313]{\includegraphics[width=.33\textwidth]{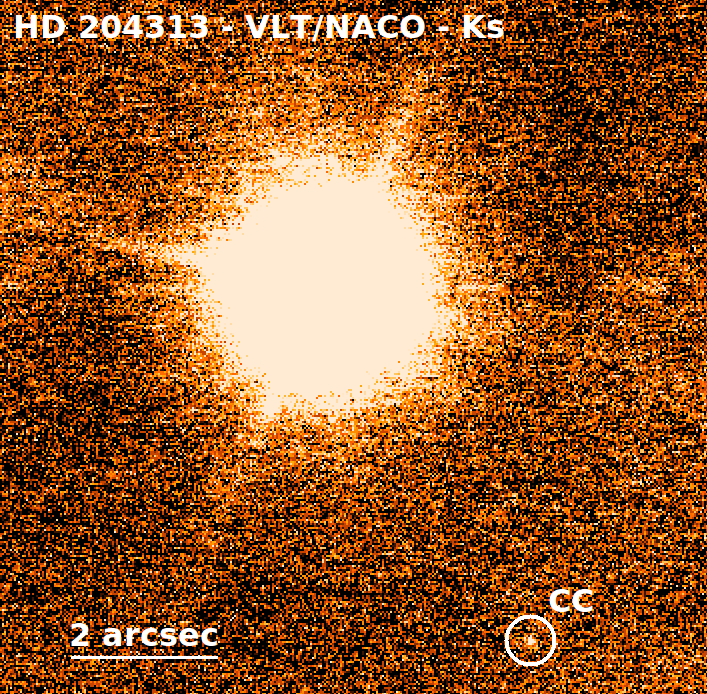}}
        
    \caption{Companion candidates detected in the VLT/NaCo archival data. Companion candidates are marked with white circles. North is up and east is to the left for all images.}
    \label{fig:companion_candidates_VLT_figure}
\end{figure*}

\begin{figure*}
        \centering
        \subfloat[GJ 176]{\includegraphics[width=.33\textwidth]{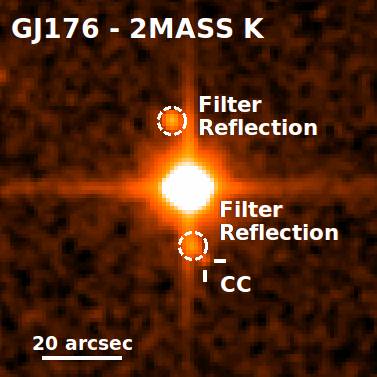}}
        \subfloat[HD 40307]{\includegraphics[width=.33\textwidth]{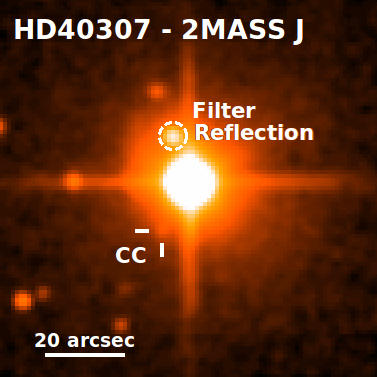}}
        \subfloat[HD 85390]{\includegraphics[width=.34\textwidth]{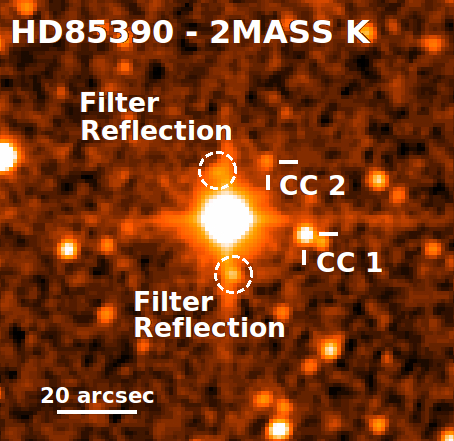}\label{hd85390_2mass}}
    \caption{Companion candidates for three systems as seen from 2MASS. The point sources in the 2MASS dashed circles are filter ghosts, while companion candidates detected in the first epoch of NaCo images are marked with white bars. In all images north is to the top and east to the left.}
    \label{fig:companion_candidates_2MASS_figure}
\end{figure*}

\begin{figure*}
        \centering
        \subfloat[]{\includegraphics[width=.5\textwidth]{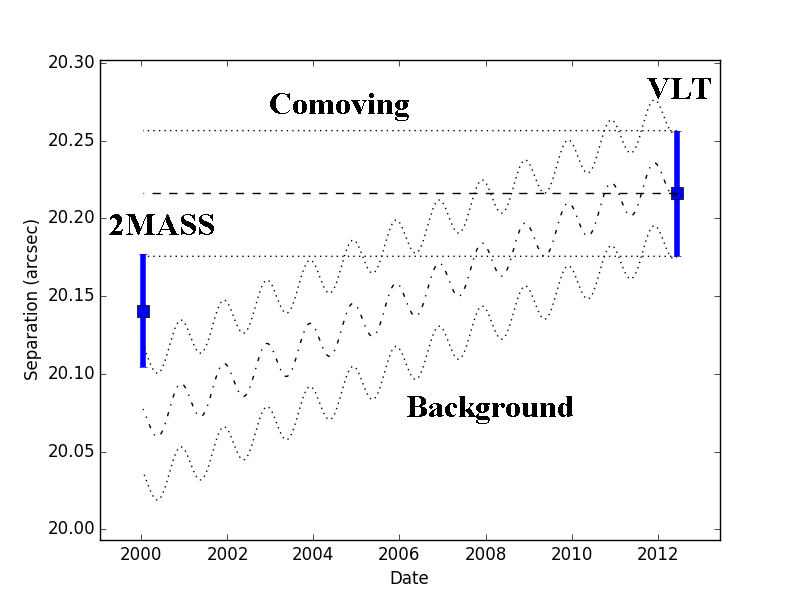}}
        \subfloat[]{\includegraphics[width=.5\textwidth]{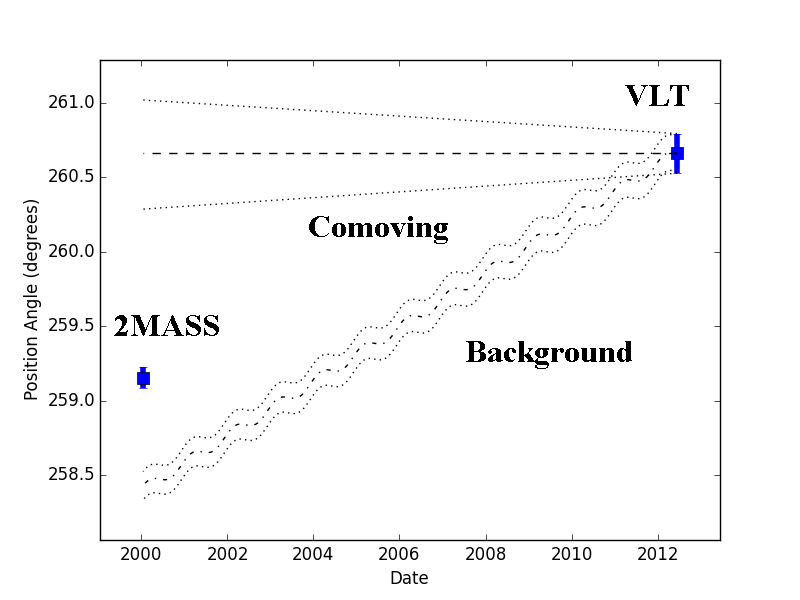}}

        \subfloat[]{\includegraphics[width=.5\textwidth]{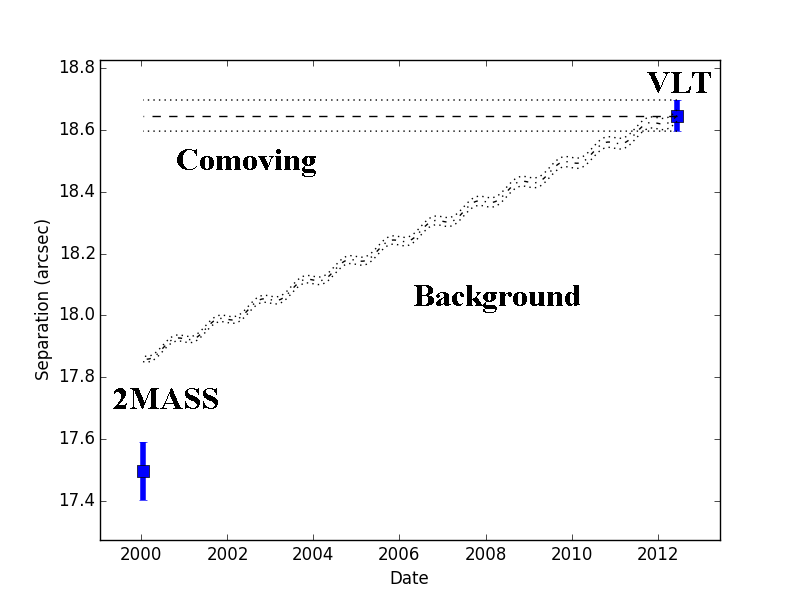}}
        \subfloat[]{\includegraphics[width=.5\textwidth]{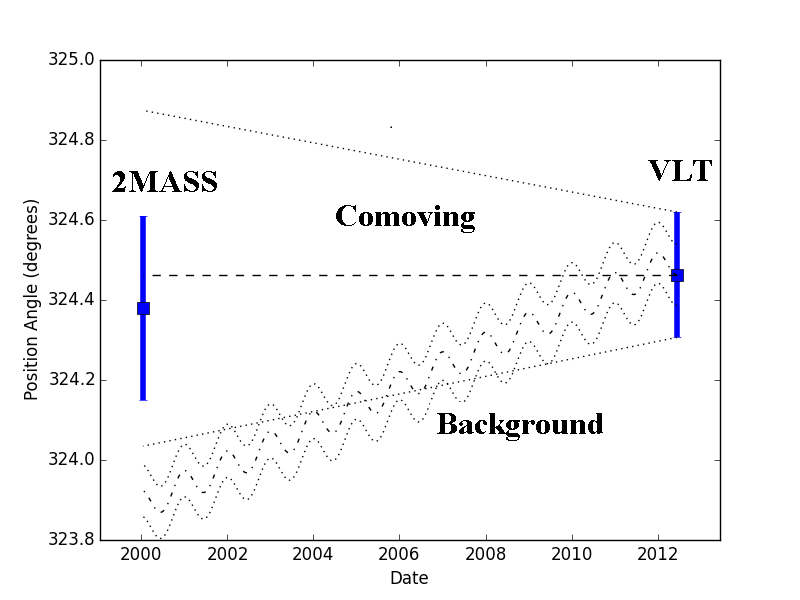}}
    \caption{Separation (left panels) and position angle (right panels) analysis for HD 85390 CC 1 (a,b) and CC 2 (c,d).}
    \label{fig:Analysis_figure_3}
\end{figure*}

In the case of HD\,85390 system we identify the two brightest companion candidates (cc1 and cc2 in Fig.~\ref{hd85390_naco}) in the 2MASS data (Fig.~\ref{hd85390_2mass}). The remaining ten candidates are presumably too close to the primary star to be resolved by 2MASS or too faint to be detected, or both. In Fig.~\ref{fig:Analysis_figure_3} we show the common proper motion plots for these two companion candidates. Both candidates clearly show differential proper motion compared with HD\,85390. For cc1 we find that the separation is in principle within 2\,$\sigma$ consistent with common proper motion and background hypothesis. 
However, in position angle we can rule out common proper motion of this companion candidate with $> 5\,\sigma$. This companion candidate is thus most likely a background star with a non-zero proper motion. The second companion candidate of HD\,85390 that we also detected in 2MASS (cc2) can also be ruled out as comoving companion based on the $> 5\,\sigma$ deviation between the NACO and 2MASS measurement in separation. We again conclude that this is most likely a background star with a non-zero proper motion. Finding multiple such background stars that are close enough to exhibit a proper motion of their own is not surprising given that HD\,85390 is located in the direction of the galactic disk.\\ 

\subsubsection{HD\,204313}
\begin{figure*}
        \centering
        \subfloat[]{\includegraphics[width=.5\textwidth]{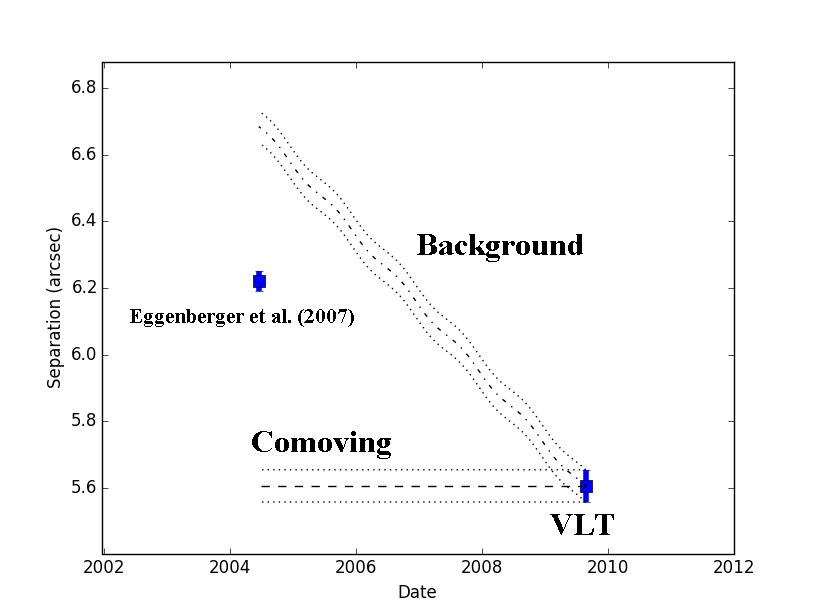}}
        \subfloat[]{\includegraphics[width=.5\textwidth]{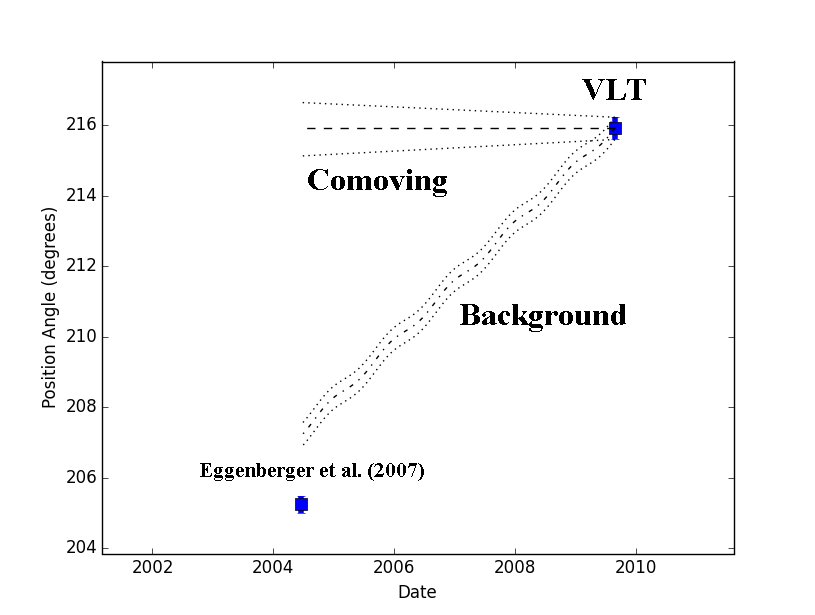}}
    \caption{Separation (left panel) and position angle (right panel) analysis for HD\,204313 CC. The 2004 data point was taken from \protect\cite{2007A&A...474..273E}.}
    \label{fig:Analysis_figure_4}
\end{figure*}

Finally, for the HD\,204313 system (Fig.~\ref{fig:Analysis_figure_4}) the two astrometric epochs are neither consistent with common proper motion nor with a non-moving background object within 3\,$\sigma$. Given that the deviation from both hypotheses is much larger than our astrometric error bars, we conclude that the most likely explanation is that the companion candidate is a background source with a non-zero proper motion. We can, however, exclude that the object is associated to the HD\,204313 system. 

\begin{table}
        \centering
        \caption{Possible companions, separation, and position angle as extracted from 2MASS}
        \label{tab:2MASS_data_table}
        \begin{tabular}{lccr}
                \hline
                System & CC & Separation (arcsec) & Position angle (deg)\\
                \hline
                GJ 176 & 1 & 19.9 ($\pm$1.4) & 186.8 ($\pm$3.1)\\
                HD 40307 & 1 & 14.24 ($\pm$0.38) & 149.8 ($\pm$1.1)\\
                HD 85390 & 1 & 20.141 ($\pm$0.037) & 259.156 ($\pm$0.071)\\
                HD 85390 & 2 & 17.498 ($\pm$0.095) & 324.38 ($\pm$0.23)\\
                \hline
        \end{tabular}
\end{table}
                
\section{Background probability of single epoch companion candidates}

To assess the background probability of the companion candidates for which we only have one observation epoch, we followed the approach of \cite{2014A&A...566A.103L}.
They estimate the probability of finding a background or foreground object within a certain separation of the host star to be

\begin{equation}
P(r,b,m_{\odot},\Delta m_{max}) = \pi r^2 \rho (b,m_{\odot},\Delta m_{max}) \ .  
\label{prob} 
\end{equation}

In this equation $r$ denotes the separation from the host star, $b$ its galactic latitude, $m_{\odot}$ is the apparent magnitude of the host star, $\Delta m_{max}$ the achieved contrast limit and finally $\rho$ is the density of background and foreground sources. We note that this equation can be interpreted only as probability for values of $r$ such that $P < 1$. For larger values of $r$ and $P$ the above equation simply gives an estimate on the expected number of background sources within the separation $r$.
To estimate $\rho$ depending on the galactic latitude, host star brightness and contrast limit we utilized the TRILEGAL\footnote{http://stev.oapd.inaf.it/cgi-bin/trilegal\_1.6} population synthesis model (\citealt{2005A&A...436..895G}). 
We used the default values as input and the initial mass function by \cite{2001ApJ...554.1274C}. 
Results are shown in Table \ref{tab:Bkgd_prb_table}. We find that the companion candidates detected around the HD\,113538 system as well as the HD\,85390 system have a high probability to be background objects. 
Conversely, the background probability of the closest companion candidate around HD\,190984 is low, making it a good candidate for follow-up observations.\\

\begin{table}
        \centering
        \caption{Background probability of single epoch companion candidates depending on separation and contrast limit. Values larger than one give the number of expected background sources rather than a probability.}
        \label{tab:Bkgd_prb_table}
        \begin{threeparttable}
        \begin{tabular}{lcccr}
                \hline
                System & CC & Sep. (arcsec) & $\Delta$\,K (mag) & P\\
                \hline
                HD\,113538 & 1 & 4.7 & 19.1 & 0.61\\
                HD\,113538 & 12 & 14.7 & 19.6 & 7.9\\
                HD\,85390 & 13 & 20\tnote{1}    & 17.5  & 13.4   \\
                HD\,190984 & 1 & 6.6 & 16.1 & 0.04\\
                HD\,190984 & 3 & 11.6 & 17.8 & 0.33\\
                \hline
        \end{tabular}
                \begin{tablenotes}
\item[1] We use the maximum separation of the outermost cc in this case since most ccs exhibit similar brightness. 
\end{tablenotes}
        \end{threeparttable}

\end{table}

\section{Detection limits}

To enable us to find solid statistics on the multiplicity rate, we determined detection limits for all our studied targets. For the image analysis, the program goes through each image and creates multiple S/N maps. First, it takes a 9x9 box around each pixel and calculates the standard deviation of the box, saving that value in that pixel as the noise value. After performing this operation on every pixel, the pipeline divides the master reduced science image by the noise map. The program also finds the center of the main star in the image and averages the boxed noise values in rings of radius 1 pixel. This average value is saved in every pixel in the ring as a new noise map. Once the ring map is created, the program again divides the reduced science image by the new noise map. Finally, the pipeline also creates a circular noise pattern by calculating the standard deviation of the pixels in rings of radius 1 pixel, saving that value in each pixel. After creating the last two S/N maps, the program uses them to calculate detection limits.\\
  The Python pipeline takes a set of separations from the central star and calculates the difference in flux between the peak value and the signal-to-noise at each separation to calculate the magnitude detection limit at a confidence of 5$\sigma$, for both the ring-box noise map and the circular noise map. It also has an interpolation routine to calculate the magnitude for a possible companion candidate star of 0.1 $M_{\sun}$, given that it would be the same age and at the same distance as the main target star. We list input distances, ages and magnitudes for our calculation in Table~\ref{tab:photometry_table}. We show the resulting detection limits for each system in Fig.~\ref{fig:detection_limits_figure}. The detection limits are calculated out to the edge of the image closest to the star, which can reach out to 14.7 arcseconds for a star imaged in the center of the detector with 27 mas pixel scale. For some systems, there is a significant difference between the ring-box and the circular magnitude detections, but these were the systems with relatively worse seeing. In the case of the Wollaston-SDI images, the combination of multiple parts of the image stacked together has a short separation cutoff, usually around 2 arcseconds. In addition, we give mass detection limits for several separations in Table~\ref{tab:detection_limits_table} using the BT-SETTL isochrones for the mass estimates (\citealt{2011ASPC..448...91A}.\\
  In the case of the HD\,40307 system we used a different approach to determine the detection limits of the archival data set. This was necessary because the data was taken in the ADI imaging mode. While this enables a superior suppression of stellar speckles and thus in general much deeper detection limits close to the bright primary star, it also introduces a self-subtraction effect to all potential companion candidates. This is due to insufficient field rotation between individual exposures, and thus a slight overlap of the PSFs of sources around the primary star. To take this effect into account we inserted fake companions into the data set. To this end, we used the unsaturated stellar PSF and scaled it to different flux levels. After insertion of the fake companions we reduced the data and create a ring S/N map as detailed earlier. We then determined the S/N of the inserted fake companions and repeat the process until we found the exact peak flux counts for a potential companion that will lead to a 5$\sigma$ detection. This is done for various separations from the primary star. For each separation we inserted four fake companions at different position angles and used the average S/N of the four sources. The resulting magnitude detection limits for HD\,40307 is shown in Fig.~\ref{adi-contrast}, while the converted mass limits are also given in Table~\ref{tab:detection_limits_table}. This procedure was not used for the HD\,98649 system, since effectively no field rotation was achieved during the observations and the data set was thus collapsed in a non-ADI way.\\

\section{Conclusions}

In this study we have analyzed archival VLT/NaCo images of 39 exoplanet host stars to study their stellar multiplicity. 
We found previously unknown companion candidates around five of these stars. The companion candidate around GJ\,176, the companion candidate around HD\,40307, and the companion candidates \#1 and \#2 around HD\,85390 were also detected in the 2MASS catalog. In addition, the companion candidate to HD\,204313 was previously detected by \cite{2007A&A...474..273E}, also with VLT/NaCo. 
We found that the companion candidates to GJ\,176 and HD\,40307 are clearly background objects, while the analysis for the two far companion candidates to HD\,85390 as well as the companion to HD\,204313 are inconclusive but also most likely explained by background objects with a non-zero proper motion. For the closer companion candidates around HD\,85390 as well as the companion candidates to HD\,113538 and HD\,190984, no suitable second observation epoch was found and follow up observations need to be performed to test if they are gravitationally bound to the primary stars. 
We used theoretical population synthesis tools to estimate the probability that these objects are unrelated background sources and found that this is indeed likely for all of these companion candidates, with the exception of the closest companion candidate to HD\,190984.
Given the ages of the systems, as well as their distances and the NaCo photometry of the companion candidates, we estimate that the bright objects around HD\,113538 and HD\,190984 would be on the border between stellar and brown dwarf, while the remaining companion candidates around HD\,85390 and HD\,113538 would be most likely brown dwarfs if indeed bound to the primary star. \\ 
Bound stellar companions may have an influence on the orbit parameters of close in planets around the primary stellar component. We found that the closest companion candidate to HD\,190984 has the highest probability of being an actual bound companion. We note that indeed the planet discovered around this star exhibits an eccentricity of 0.57 (\citealt{2010AandA...512A..47S}).
Kozai-Lidov interactions with a potential stellar companion would be an explanation for this high eccentricity.\\
Given the multiplicity rate of 12\,\% recovered by \cite{2012A&A...542A..92R}, we would have expected five new stellar companions in our sample of 39 stars. However, our sample is likely biased toward non-detections, since newly detected stellar companions probably have a higher chance of being published than non-detections. It is in principle still possible (but unlikely given our analysis) that we found five or more new companions pending follow-up observations of the mentioned companion candidates.\\
We calculated detection limits for all stars in our sample. Due to the nature of our study, our sample is inhomogeneous in target properties as well as observation modes. This is reflected by the individual detection limits.  
On average we found a 5\,$\sigma$ contrast limit of $\Delta$\,m $\sim$6.5\,mag for 1\,arcsec, and $\Delta$\,m $>$8\,mag outside of 2.5\,arcsec. This enables us to exclude $\sim$ 0.1\,M$_\odot$ companions outside of 2.5\,arcsec around the majority of our target stars.
Comparing to the literature we find that our detection limits are similar to those achieved in several recent studies, carried out with smaller (2\,m class) telescopes (\citealt{2014ApJ...791...35L}, \citealt{2016MNRAS.457.2173G}, \citealt{2017AJ....153...66Z}). 
However, our results are worse than previous VLT/NACO studies, for example, \cite{2007A&A...474..273E} find a K-band contrast limit of $\sim$8\,mag at separations larger than 0.65\,arcsec.
We attribute this to the mentioned inhomogeneity of our data in terms of observation modes, but also observing conditions.\\
Our study shows that archival data can contribute to determine the true stellar multiplicity rate of exoplanet host stars. While our study can be considered complete for all known exoplanet host stars observed with VLT/NaCo, there is of course a wealth of other instruments available that we did not consider. In particular, the Gemini, Keck and Subaru archives should be scanned in a similar way to ours.

\begin{acknowledgements}
     The authors would like to thank John Tobin and Silvia Toonen for organizing the LEAPS (Leiden/ESA Astrophysics Program for Summer Students) program in the context of which this study was executed, as well as Christoph Keller for organizing the necessary funding. This research has made use of the SIMBAD database, operated at CDS, Strasbourg, France. This research made use of the VizieR catalog access tool, CDS, Strasbourg, France. This research has made use of NASA's Astrophysics Data System. This research made use of matplotlib, a Python library for publication quality graphics \citep{Hunter:2007}. CG wishes to thank Donna Keeley for some language editing of the manuscript. JD would like to thank Jerry Dietrich for some assistance with Python programming and pipeline management.
\end{acknowledgements}

%
%

\bibliographystyle{aa} 
\bibliography{leaps.bib} 

\begin{appendix} 

\section{Data tables}

\begin{sidewaystable*}
	\centering
	\caption{The system, number of images, exposure time for one image, filter, mode, and program ID in which each system was imaged}
	\label{tab:VLT_archives_table}
	\begin{threeparttable}
	\begin{tabular}{lcccccccr}
                \hline
                System & \# Images & Exp. time (s) & Observation date & Filter & Mode & Pixel scale (arcsec) & True north ($^\circ$) & Program ID \\
                \hline
                GJ 176 & 630 & 0.4 & Sep 15 2007 & NB-2.17 & Jitter & 0.02715 & -0.74 & 079.C-0216(A)\\
                GJ 436 & 814 & 0.8 & Feb 25 2007 & H; Wollaston-SDI & SDI Jitter & 0.01732 & -0.74 & 278.C-5013(B)\\
                GJ 581 & 36 & 5.0 & Jun 29, 2008 & Ks & Jitter & 0.02715 & -0.74 & 081.C-0600(A)\\
                GJ 849 & 165 & 0.5 & Nov 17, 18, \& 20; 2006 & H; Wollaston-SDI & SDI Jitter & 0.01732 & -0.74 & 278.C-5013(A)\\
                GJ 876 & 759 & 4.0 & May 29 2005 & H; Wollaston-SDI & SDI Jitter & 0.01732 & -0.32 & 075.C-0357(A)\\
                HD 4113 & 30 & 5.0 & Aug 20 2008 & Ks & Jitter & 0.02715 & -1.01 & 081.C-0653(A)\\
                HD 10647 & 300 & 1.0 & Aug 25 2006 & Ks & Jitter & 0.01327 & -0.78 & 077.C-0444(A)\\
                HD 11506 & 200 & 1.0 & Aug 24 2007 & ND Ks & Jitter & 0.02715 & -0.74 & 079.C-0689(B)\\
                HD 11964 & 510 & 0.35 & Aug 27 2009 & ND Ks & Jitter & 0.02715 & -1.01 & 083.C-0151(A)\\
                HD 20782 & 45 & 2.0 & Aug 24 2007 & ND Ks & Jitter & 0.02715 & -0.74 & 079.C-0689(B)\tnote{1}\\
                HD 20782 & 180 & 1.0 & Jun 28 2008 & ND Ks & Jitter & 0.02715 & -0.74 & 081.C-0600(A)\tnote{1}\\
                HD 20782 & 300 & 0.5 & Sep 27 2009 & ND Ks & Jitter & 0.02715 & -0.74 & 083.C-0599(A)\tnote{1}\\
                HD 21693 & 70 & 12.0 & Dec 6 2005 & H; Wollaston-SDI & SDI Jitter & 0.01732 & -0.6 & 076.C-0762(A)\\
                HD 23127 & 60 & 5.0 & Aug 25 2007& ND Ks & Jitter & 0.02715 & -0.74 & 079.C-0689(B)\\
                HD 30562 & 600 & 0.35 & Sep 27 2009 & ND Ks & Jitter & 0.02715 & -1.01 & 083.C-0599(A)\\
                HD 33283 & 60 & 5.0 & Aug 25 2007& ND Ks & Jitter & 0.02715 & -0.74 & 079.C-0689(B)\\
                HD 40307 & 3780 & 0.5 & Dec 31 2011 & Ks & ADI Cube & 0.02715 &       -       & 088.C-0832(A)\\
                HD 47186 & 50 & 2.0 & Mar 30 2010 & ND Ks & Jitter & 0.02715 & -1.01 & 385.C-0390(A)\\
                HD 60532 & 180 & 0.35 & Sep 26 2010 & ND Ks & Jitter & 0.02715 & -1.01 & 085.C-0277(B)\\
                HD 69830 & 7350 & 0.3453 & Oct 27 2010 & H & Jitter & 0.02715 & -1.01 & 082.C-0518(B)\\
                HD 85390 & 1800 & 0.11 & Jun 14 2012 & Ks & Jitter Cube & 0.02715 & -1.01 & 089.C-0641(A)\\
                HD 85512 & 1638 & 0.3454 & Jan 24 2013 & Ks & Jitter Cube & 0.01327 & -1.26 & 090.C-0125(A)\\ 
                HD 98649 & 12700 & 55.5 & Feb 8 2009 & Ks & ADI Jitter Cube\tnote{2} & 0.02715 & -1.01 & 082.C-0184(B)\\
                HD 113538 & 7000 & 0.347 & May 12 2008 & Ks & Jitter & 0.02715 & -0.74 & 081.D-0012(A)\\
                HD 114613 & 2880 & 1.0 & Feb 02 2004 & H; Wollaston-SDI & SDI Jitter & 0.0172 & -0.22 & 60.A-9026(A)\\
                HD 142022 & 100 & 2.0 & Jun 29 2008 & ND Ks & Jitter & 0.02715 & -1.01 & 081.C-0600(A)\tnote{1}\\
                HD 142022 & 90 & 1.0 & Jun 18 2010 & ND Ks & Jitter & 0.02715 & -1.01 & 385.C-0390(A)\tnote{1}\\
                HD 147873 & 96 & 1.0 & May 27 2005 & NB 2.17 & Jitter & 0.01327 & -0.32 & 075.C-0668(A)\\
                HD 154857 & 25 & 0.5 & Aug 24 2007 & ND Ks & Jitter & 0.02715 & -0.74 & 079.C-0689(B)\tnote{1}\\
                HD 154857 & 380 & 0.5 & Sep 26 2009 & ND Ks & Jitter & 0.02715 & -0.74 & 083.C-0599(A)\tnote{1}\\
                HD 159868 & 400 & 0.5 & Jun 29 2008 & ND Ks & Jitter & 0.02715 & -0.74 & 081.C-0600(A)\\
                HD 168443 & 570 & 0.35 & Aug 24 2007 & ND Ks & Jitter & 0.02715 & -1.01 & 079.C-0689(B)\tnote{1}\\
                HD 168443 & 480 & 0.35 & Sep 26 2009 & ND Ks & Jitter & 0.02715 & -1.01 & 083.C-0599(A)\tnote{1}\\
                HD 171238 & 110 & 20.0 & Aug 02 2006 & H; Wollaston-SDI & SDI Jitter & 0.01732 & -0.78 & 077.C-0293(A)\\
                HD 175167 & 1200 & 0.5 & Jul 23 2012 & Ks & Jitter Cube & 0.02715 & -1.01 & 089.C-0638(A)\\
                HD 183263 & 54 & 1.5 & Aug 25 2007 & ND Ks & Jitter & 0.02715 & -0.74 & 079.C-0689(B)\tnote{1}\\
                HD 183263 & 96 & 1.5 & Sep 28 2009 & ND Ks & Jitter & 0.02715 & -0.74 & 083.C-0599(A)\tnote{1}\\
                HD 183263 & 108 & 1.5 & Mar 20 2010 & ND Ks & Jitter & 0.02715 & -0.74 & 084.C-0443(A)\tnote{1}\\
                HD 187085 & 100 & 0.5 & Sep 27 2009 & ND Ks & Jitter & 0.02715 & -0.74 & 083.C-0599(A)\\
                HD 190647 & 100 & 1.0 & Sep 27 2009 & ND Ks & Jitter & 0.02715 & -0.74 & 083.C-0599(A)\\
                HD 190984 & 1200 & 0.5 & Jul 23 2012 & Ks & Jitter Cube & 0.02715 & -1.01 & 089.C-0638(A)\\
                HD 204313 & 770 & 0.5 & Aug 27 2009 & NB-2.17 & Jitter & 0.02715 & -0.74 & 083.C-0151(A)\\
                HD 208487 & 200 & 1.0 & Aug 24 2007 & ND Ks & Jitter & 0.02715 & -0.74 & 079.C-0689(B)\\
                HD 210702 & 600 & 0.35 & Jun 29 2008 & ND Ks & Jitter & 0.02715 & -0.74 & 081.C-0600(A)\\
                HD 219077 & 1200 & 1.6 & Jul 24 2006 & H; Wollaston-SDI & SDI Jitter & 0.01732 & -0.78 & 077.C-0293(A)\\
                HD 221287 & 120 & 5.0 & Aug 25 2007 & ND Ks & Jitter & 0.02715 & -0.74 & 079.C-0689(B)\\
                \hline
        \end{tabular}
		\begin{tablenotes}
\item[1] Combined into one set of observations
\item[2] Due to the declination of the target and the time of the observation, there was effectively no field rotation achieved.
\item[3] Used for proper motion analysis
\end{tablenotes}
\end{threeparttable}
\end{sidewaystable*}

\begin{table*}
        \centering
        \caption{Possible companions analyzed, separation, and position angle as extracted from the VLT/NaCo data.}
        \label{tab:VLT_data_table}
        \begin{tabular}{lccrrr}
                \hline
                System & CC & Epoch & Separation (arcsec) & Position angle (deg)& $\Delta$K (mag)  \\
                \hline
                GJ 176 & 1 & Sep 15 2007 & 13.603 ($\pm$0.046) & 215.66 ($\pm$0.20) & 9.34 ($\pm$0.29) \\
                HD 40307 & 1 & Dec 31 2011 & 13.350 ($\pm$0.051) & 146.31 ($\pm$0.21) & 8.54 ($\pm$0.93) \\
                HD 85390 & 1 & Jun 14 2012 & 20.216 ($\pm$0.037) & 260.66 ($\pm$0.17) & 6.70 ($\pm$0.03) \\
                HD 85390 & 2 & Jun 14 2012 & 18.647 ($\pm$0.038) & 324.46 ($\pm$0.17) & 9.23 ($\pm$0.08) \\
                HD 85390 & 3 & Jun 14 2012 & 10.402 ($\pm$0.021) & 275.62 ($\pm$0.17) &  11.14 ($\pm$0.22) \\
                HD 85390 & 4 & Jun 14 2012 & 4.499 ($\pm$0.015) & 352.43 ($\pm$0.18) & 10.71 ($\pm$0.17) \\
                HD 85390 & 5 & Jun 14 2012 & 20.758 ($\pm$0.040) & 345.41 ($\pm$0.17) & 10.88 ($\pm$0.18) \\
                HD 85390 & 6 & Jun 14 2012 & 18.853 ($\pm$0.036) & 186.80 ($\pm$0.17) & 10.87 ($\pm$0.18) \\
                HD 85390 & 7 & Jun 14 2012 & 8.740 ($\pm$0.018) & 148.05 ($\pm$0.17) & 11.47 ($\pm$0.25) \\
                HD 85390 & 8 & Jun 14 2012 & 9.827 ($\pm$0.019) & 144.44 ($\pm$0.17) & 12.45 ($\pm$0.45) \\
                HD 85390 & 9 & Jun 14 2012 & 6.184 ($\pm$0.014) & 109.38 ($\pm$0.17) & 11.74 ($\pm$0.28) \\
                HD 85390 & 10 & Jun 14 2012 & 8.717 ($\pm$0.017) & 109.29 ($\pm$0.17) & 10.74 ($\pm$0.17) \\
                HD 85390 & 11 & Jun 14 2012 & 10.295 ($\pm$0.020) & 100.88 ($\pm$0.17) & 10.48 ($\pm$0.15) \\
                HD 85390 & 12 & Jun 14 2012 & 15.463 ($\pm$0.031) & 95.75 ($\pm$0.17) & 10.83 ($\pm$0.18) \\
                HD 85390 & 13 & Jun 14 2012 & 6.940 ($\pm$0.015) & 252.72 ($\pm$0.17) & 11.21 ($\pm$0.22) \\
                HD 113538 & 1 & May 12 2008 & 4.649 ($\pm$0.291) & 118.56 ($\pm$0.18) & 13.46 ($\pm$0.72) \\
                HD 113538 & 2 & May 12 2008 & 7.052 ($\pm$0.018) & 226.14 ($\pm$0.17) & 12.46 ($\pm$0.34) \\
                HD 113538 & 3 & May 12 2008 & 7.044 ($\pm$0.188) & 305.48 ($\pm$0.17) & 13.55 ($\pm$0.60) \\
                HD 113538 & 4 & May 12 2008 & 7.513 ($\pm$0.437) & 70.90 ($\pm$0.18) & 14.12 ($\pm$0.80) \\
                HD 113538 & 5 & May 12 2008 & 8.972 ($\pm$0.018) & 132.90 ($\pm$0.17) & 11.78 ($\pm$0.24) \\
                HD 113538 & 6 & May 12 2008 & 10.285 ($\pm$0.021) & 98.88 ($\pm$0.17) & 11.32 ($\pm$0.20) \\
                HD 113538 & 7 & May 12 2008 & 10.606 ($\pm$0.026) & 106.78 ($\pm$0.17) & 12.71 ($\pm$0.38) \\
                HD 113538 & 8 & May 12 2008 & 10.950 ($\pm$0.028) & 153.71 ($\pm$0.17) & 12.34 ($\pm$0.32) \\
                HD 113538 & 9 & May 12 2008 & 12.073 ($\pm$0.023) & 188.82 ($\pm$0.17) & 10.88 ($\pm$0.16) \\
                HD 113538 & 10 & May 12 2008 & 14.263 ($\pm$0.030) & 148.76 ($\pm$0.17) & 11.93 ($\pm$0.26) \\
                HD 113538 & 11 & May 12 2008 & 12.304 ($\pm$0.034) & 81.48 ($\pm$0.17) & 14.49 ($\pm$0.84) \\
                HD 113538 & 12 & May 12 2008 & 14.718 ($\pm$0.032) & 205.73 ($\pm$0.17) & 13.95 ($\pm$0.66) \\
                HD 190984 & 1 & Jul 23 2012 & 6.621 ($\pm$0.020) & 202.83 ($\pm$0.17) & 8.83 ($\pm$0.09) \\
                HD 190984 & 2 & Jul 23 2012 & 7.486 ($\pm$0.020) & 128.74 ($\pm$0.17) & 9.46 ($\pm$0.12) \\
                HD 190984 & 3 & Jul 23 2012 & 11.636 ($\pm$0.027) & 31.66 ($\pm$0.17) & 10.87 ($\pm$0.23) \\
                HD 204313 & 1 & Aug 27 2009 & 5.606 ($\pm$0.039) & 215.91 ($\pm$0.33) & 9.74 ($\pm$0.40) \\
                \hline
        \end{tabular}
\end{table*}

\begin{table*}[!h]
        \centering
        \caption{Apparent magnitudes, distance, and age for each system, where the apparent magnitudes correspond to the 2MASS filter closest to the filter in which the VLT NaCo observations were made (H or K)}
        \label{tab:photometry_table}
        \begin{tabular}{lccccccr}
                \hline
                Primary & App. Mag & Filter &  Ref. & Distance (pc) & Ref. & Age & Ref.\\
                \hline
                GJ 176 & 5.607 ($\pm$0.034) & K & (1) & 9.27 ($\pm$0.24) & (2) & 2.3$^{+6.4}_{-1.7}$ & (3)\\
                GJ 436 & 6.319 ($\pm$0.023) & H & (1) & 10.13 ($\pm$0.24) & (2) & 9.47($\pm$0.57) & (3)\\
                GJ 581 & 5.837 ($\pm$0.023) & H & (1) & 6.21 ($\pm$0.10) & (2) & 9.44 ($\pm$0.58) & (3)\\
                GJ 849 & 5.899 ($\pm$0.044) & K & (1) & 115.05 ($\pm$7.36) & (2) & 9.4 ($\pm$0.6) & (3)\\
                GJ 876 & 5.349 ($\pm$0.049) & H & (1) & 4.69 ($\pm$0.47) & (2) & 9.51 ($\pm$0.58) & (3)\\
                HD 4113 & 6.345 ($\pm$0.021) & K & (1) & 44.0 ($\pm$1.7) & (2) & 8.9 ($\pm$2.5) & (2)\\
                HD 10647 & 4.340 ($\pm$0.276) & K & (1) & 17.43 ($\pm$0.07) & (4) & 3.2 ($\pm$1.2) & (5)\\
                HD 11506 & 6.168 ($\pm$0.017) & K & (1) & 51.7 ($\pm$1.6) & (4) & 1.6 ($\pm$0.9) & (5)\\
                HD 11964 & 4.491 ($\pm$0.020) & K & (1) & 32.84 ($\pm$0.66) & (4) & 8.5 ($\pm$0.5) & (6)\\
                HD 20782 & 5.827 ($\pm$0.016) & K & (1) & 35.5 ($\pm$0.8) & (4) & 5.4 ($\pm$1.3) & (6)\\
                HD 21693 & 6.288 ($\pm$0.024) & H & (1) & 32.37 ($\pm$0.52) & (2) & 5.9 ($\pm$4.3) & (7)\\
                HD 23127 & 7.087 ($\pm$0.018) & K & (1) & 98.2 ($\pm$6.5) & (2) & 4.8 ($\pm$0.6) & (5)\\
                HD 30562 & 4.310 ($\pm$0.049) & K & (8) & 26.43 ($\pm$0.24) & (9) & 4.4 ($\pm$0.6) & (5)\\
                HD 33283 & 6.646 ($\pm$0.033) & K & (1) & 94.2 ($\pm$5.7) & (4) & 3.6 ($\pm$0.6) & (5)\\
                HD 40307 & 4.793 ($\pm$0.016) & K & (1) & 13.00 ($\pm$0.07) & (2) & 7.0 ($\pm$4.2) & (5)\\
                HD 47186 & 6.005 ($\pm$0.027) & K & (1) & 39.6 ($\pm$1.0) & (2) & 5.5 ($\pm$0.6) & (5)\\
                HD 60532 & 3.355 ($\pm$0.286) & K & (1) & 25.3 ($\pm$0.2) & (2) & 3.0 ($\pm$0.2) & (6)\\
                HD 69830 & 4.118 ($\pm$0.094) & H & (10) & 12.49 ($\pm$0.05) & (2) & 10.4 ($\pm$2.5) & (6)\\
                HD 85390 & 6.491 ($\pm$0.023) & K & (1) & 32.36 ($\pm$0.65) & (2) & 5.6 ($\pm$3.7) & (5)\\
                HD 85512 & 4.717 ($\pm$0.021) & K & (1) & 11.16 ($\pm$0.08) & (2) & 8.2 ($\pm$3.0) & (6)\\
                HD 98649 & 6.419 ($\pm$0.021) & K & (1) & 41.5 ($\pm$1.4) & (9) & 4.9 ($\pm$3.5) & (7)\\
                HD 113538 & 5.637 ($\pm$0.024) & K & (1) & 15.86 ($\pm$0.35) & (2) & 4.3 ($\pm$4.0) & (11)\\
                HD 114613 & 3.346 ($\pm$0.206) & H & (1) & 20.67 ($\pm$0.12) & (2) & 5.0 ($\pm$0.1) & (6)\\
                HD 142022 & 5.964 ($\pm$0.027) & K & (1) & 34.31 ($\pm$0.69) & (2) & 5.2 ($\pm$3.5) & (12)\\
                HD 147873 & 6.581 ($\pm$0.029) & K & (1) & 103.59 ($\pm$10.77) & (2) & 2.88 ($\pm$0.64) & (7)\\
                HD 154857 & 5.509 ($\pm$0.018) & K & (1) & 64.07 ($\pm$2.95) & (2) & 5.8 ($\pm$0.5) & (6)\\
                HD 159868 & 5.535 ($\pm$0.024) & K & (1) & 52.7 ($\pm$3.0) & (13) & 6.1 ($\pm$0.4) & (5)\\
                HD 168443 & 5.211 ($\pm$0.015) & K & (1) & 37.4 ($\pm$0.97) & (2) & 10.0 ($\pm$0.3) & (6)\\
                HD 171238 & 6.868 ($\pm$0.023) & H & (1) & 50.3 ($\pm$2.9) & (14) & 2.5 ($\pm$1.6) & (5)\\
                HD 175167 & 6.288 ($\pm$0.018) & K & (1) & 67.09 ($\pm$3.35) & (2) & 8.35 ($\pm$1.35) & (15)\\
                HD 183263 & 6.422 ($\pm$0.018) & K & (1) & 54.92 ($\pm$2.80) & (2) & 4.5 ($\pm$ 0.8) & (6)\\
                HD 187085 & 5.876 ($\pm$0.024) & K & (1) & 44.0 ($\pm$1.3) & (4) & 3.6 ($\pm$0.3) & (5)\\
                HD 190647 & 6.161 ($\pm$0.038) & K & (1) & 57.1 ($\pm$2.6) & (2) & 8.3 ($\pm$0.5) & (5)\\
                HD 190984 & 7.319 ($\pm$0.016) & K & (1) & 103 ($\pm$21) & (16) & 7.8 ($\pm$1.4) & (17)\\
                HD 204313 & 6.459 ($\pm$0.018) & K & (1) & 47.3 ($\pm$1.4) & (2) & 5.5 ($\pm$0.8) & (5)\\
                HD 208487 & 6.159 ($\pm$0.034) & K & (8) & 45.8 ($\pm$1.4) & (2) & 3.1 ($\pm$0.6) & (5)\\
                HD 210702 & 3.98 ($\pm$0.29) & K & (1) & 55.0 ($\pm$1.1) & (18) & 2.1 ($\pm$0.1) & (5)\\
                HD 219077 & 4.12 ($\pm$0.14) & H & (1) & 29.35 ($\pm$0.30) & (9) & 8.9 ($\pm$0.3) & (9)\\
                HD 221287 & 6.569 ($\pm$0.020) & K & (1) & 55.2 ($\pm$1.9) & (2) & 0.8 ($\pm$0.5) & (5)\\
                \hline
        \end{tabular}
        \tablebib{(1)~\citet{2006AJ....131.1163S};
        (2) \cite{2012yCat.5137....0A}; (3) \cite{2015ApJ...804...64M}; (4) \cite{2012MNRAS.427..343M};
        (5) \cite{2015AandA...575A..18B}; (6) \cite{2016AandA...585A...5B}; (7) \cite{2013AandA...551L...8P};
        (8) \cite{2009ApJ...703L..72T}; (9) \cite{2011arXiv1109.2497M}; (10) \cite{2012yCat.2311....0C};
        (11) \cite{2015AandA...576A..48M}; (12) \cite{2013AandA...555A.150T}; (13) \cite{2012ApJ...753..169W};
        (14) \cite{2010AandA...511A..45S}; (15) \cite{2001AandA...377..911F}; (16) \cite{2010AandA...512A..47S};
        (17) \cite{2009AandA...501..941H}; (18) \cite{2012PASJ...64..135S}.}
\end{table*}

\begin{table*}
        \centering
        \caption{Main systems and mass limit (in solar masses) at intervals (in arcseconds) of 0.5, 1, 2.5, 5, and 10 (if possible).}
        \label{tab:detection_limits_table}
        \begin{threeparttable}
        \begin{tabular}{lccccc}
                \hline
                System & Mass limit at 0.5'' & Mass limit at 1'' & Mass limit at 2.5'' & Mass limit at 5'' & Mass limit at 10'' \\
                \hline
                GJ 176  &        0.065$^{+0.008}_{-0.027}$      &        0.045$^{+0.026}_{-0.021}$       &        0.036$^{+0.029}_{-0.018}$      &        0.036$^{+0.029}_{-0.018}$       &        0.036$^{+0.029}_{-0.018}$      \\
                GJ 436  &        0.07373 ($\pm$0.00015)         &        0.0638$^{+0.0042}_{-0.0006}$    &        X\tnote{1}     &        X\tnote{1}         &        X\tnote{1}     \\
                GJ 581  &        0.0714 ($\pm$0.0004)   &        0.0679$^{+0.0027}_{-0.0014}$         &        0.0654$^{+0.0047}_{-0.0007}$   &        0.0652$^{+0.0048}_{-0.0007}$         &        0.0624$^{+0.0028}_{-0.0005}$   \\
                GJ 849  &        0.353$^{+0.020}_{0.019}$       &        0.1398$^{+0.0053}_{-0.0070}$    &       X\tnote{1}      &       X\tnote{1}         &        X\tnote{1}     \\
                GJ 876  &        0.07351$^{+0.00039}_{-0.00042}$        &        0.0642$^{+0.0059}_{-0.0009}$    &        0.0529$^{+0.0015}_{-0.0025}$   &        0.0653$^{+0.0050}_{-0.0012}$    &        X\tnote{1}     \\
                HD 4113         &        0.306 $^{+0.014}_{-0.011}$     &        0.1046$^{+0.0018}_{-0.0019}$    &        0.07965$^{+0.00032}_{-0.00031}$         &        0.07958$^{+0.00032}_{-0.00031}$        &       0.07813 ($\pm$0.00031)         \\
                HD 10647        &        0.0842$^{+0.039}_{-0.0040}$    &        0.0758$^{+0.0019}_{-0.0031}$    &        0.0754$^{+0.0030}_{-0.0031}$   &        0.0717$^{+0.0020}_{-0.0052}$    &X\tnote{1}     \\
                HD 11506        &        0.1963$^{+0.0069}_{-0.0065}$   &        0.0992$^{+0.0019}_{-0.0025}$    &        0.0891$^{+0.0012}_{-0.0033}$   &        0.0899$^{+0.0018}_{-0.0029}$    &        0.0843$^{+0.0015}_{-0.0054}$   \\
                HD 11964        &        0.2070$^{+0.0045}_{-0.0044}$   &        0.1096$^{+0.0016}_{-0.0012}$    &        0.0884 ($\pm$0.0006)   &        0.0877 ($\pm$0.0006)    &        0.0828 ($\pm$0.0006)   \\
                HD 20782        &        0.3601 ($\pm$0.0090)   &        0.1277 ($\pm$0.0017)    &        0.0905 ($\pm$0.0008)   &        0.0900$^{+0.0010}_{0.0005}$         &        0.0855 ($\pm$0.0005)   \\
                HD 21693        &        0.0903$^{+0.0011}_{-0.0008}$   &        0.0782$^{+0.0002}_{-0.0024}$    &       X\tnote{1}      &        X\tnote{1}         &        X\tnote{1}     \\
                HD 23127        &        0.545 ($\pm$0.024)     &        0.279 ($\pm$0.020)      &        0.175$^{+0.013}_{-0.011}$      &        0.173$^{+0.013}_{-0.011}$       &        0.150$^{+0.011}_{-0.008}$      \\
                HD 30562        &        0.1931 ($\pm$0.0051)   &        0.08142 ($\pm$0.00073)  &        0.07597$^{+0.00038}_{-0.00044}$        &        0.07574$^{+0.00038}_{-0.00045}$         &        0.07449$^{+0.00031}_{-0.00041}$         \\
                HD 33283        &        0.435 ($\pm$0.024)     &        0.180$^{+0.013}_{-0.042}$       &        0.1274$^{+0.0070}_{-0.0060}$   &        0.1317$^{+0.0078}_{-0.0064}$    &        0.1169 ($\pm$0.0060)   \\
                HD 40307        &        0.1678 ($\pm$0.0021)   &        0.07918 ($\pm$0.00015)  &        0.07053 ($\pm$0.00016)         &        0.06771 ($\pm$0.00015)  &        0.06663 ($\pm$0.00015)         \\
                HD 47186        &        0.1082$^{+0.0020}_{-0.0019}$   &        0.084$^{+0.001}_{-0.028}$       &        0.0777$^{+0.0035}_{-0.0037}$   &        0.0775$^{+0.0035}_{-0.0037}$    &        0.0751$^{+0.0038}_{-0.0039}$   \\
                HD 60532        &        0.6601$^{+0.0511}_{-0.0496}$   &        0.2371$^{+0.0375}_{-0.0326}$    &        0.0816$^{+0.0040}_{-0.0027}$   &        0.0797$^{+0.0037}_{0.0019}$     &        0.0791$^{+0.0029}_{-0.0019}$   \\
                HD 69830        &        0.5272$^{+0.0197}_{-0.0182}$   &        0.4657$^{+0.0194}_{-0.0185}$    &        0.1859$^{+0.0101}_{-0.0095}$   &        0.2082$^{+0.0125}_{-0.0111}$    &        0.1688$^{+0.0086}_{-0.008}$    \\
                HD 85390        &        0.2461$^{+0.0107}_{-0.0082}$   &        0.0964$^{+0.0012}_{-0.0013}$    &        0.0711$^{+0.0013}_{-0.0094}$   &        0.0647$^{+0.0058}_{-0.0178}$    &        0.0542$^{+0.0079}_{-0.0211}$   \\
                HD 85512        &        0.09473$^{+0.00072}_{-0.00073}$         &        0.07358$^{+0.0028}_{-0.0090}$  &        0.0671$^{+0.0003}_{-0.0264}$         &        0.0588$^{+0.0028}_{-0.0079}$   &        X\tnote{1}     \\
                HD 98649        &        0.0689$^{+0.0024}_{-0.0193}$   &        0.0635$^{+0.0049}_{-0.0240}$    &        0.0601$^{+0.0048}_{-0.0259}$   &        0.0601$^{+0.0048}_{-0.0259}$    &        X\tnote{1}     \\
                HD 113538       &        0.1095$^{+0.0038}_{-0.0088}$   &        0.0768$^{+0.0005}_{-0.0332}$    &        0.0613$^{+0.0058}_{-0.0454}$   &        0.048$^{+0.0115}_{-0.0355}$     &        0.0458$^{+0.0115}_{-0.0341}$   \\
                HD 114613       &        0.483$^{+0.034}_{-0.037}$      &        0.1139$^{+0.0092}_{-0.0079}$    &       X\tnote{1}      &        X\tnote{1}         &       X\tnote{1}      \\
                HD 142022       &        0.1762$^{+0.0049}_{-0.0042}$   &        0.0916 ($\pm$0.0012)    &        0.0794$^{+0.0005}_{-0.001}$    &        0.0795$^{+0.0005}_{-0.0009}$    &        0.0777$^{+0.0005}_{-0.0021}$   \\
                HD 147873       &        0.1492$^{+0.0099}_{-0.0080}$   &        0.1041$^{+0.0043}_{-0.0080}$    &        0.0995$^{+0.0041}_{-0.0577}$   &        0.0922$^{+0.0035}_{-0.0471}$    &        X\tnote{1}     \\
                HD 154857       &        0.665$^{+0.0116}_{-0.0115}$    &        0.2903$^{+0.0093}_{-0.0092}$    &        0.1708$^{+0.0049}_{-0.0048}$   &        0.1713$^{+0.0051}_{-0.0048}$    &        0.1471$^{+0.004}_{-0.0038}$    \\
                HD 159868       &        0.1619 ($\pm$0.0094)   &        0.0921$^{+0.0031}_{-0.0026}$    &        0.0850 ($\pm$0.0018)   &        0.0845 ($\pm$0.0017)    &        0.0805$^{+0.0017}_{-0.0009}$   \\
                HD 168443       &        0.2882 ($\pm$0.0057)   &        0.1177 ($\pm$0.0018)    &        0.0931 ($\pm$0.001)    &        0.093 ($\pm$0.001)         &        0.0869 ($\pm$0.0006)   \\
                HD 171238       &        0.1436 ($\pm$0.0047)   &        0.0832$^{+0.0012}_{-0.0042}$    &        X\tnote{1}     &        X\tnote{1}         &        X\tnote{1}     \\
                HD 175167       &        0.4121$^{+0.0129}_{-0.0127}$   &        0.1572$^{+0.0052}_{-0.0051}$    &        0.0753 ($\pm$0.0004)   &        0.0725$^{+0.0004}_{-0.0005}$    &        0.0721$^{+0.0004}_{-0.0005}$   \\
                HD 183263       &        0.3384$^{+0.0154}_{-0.2965}$   &        0.131$^{+0.004}_{-0.0033}$      &        0.1056 ($\pm$0.0021)   &        0.1063$\pm$0.0021       &        0.0967 ($\pm$0.0017)   \\
                HD 187085       &        0.2146 ($\pm$0.0077)   &        0.0969 ($\pm$0.0017)    &        0.0897$^{+0.0014}_{-0.0009}$   &        0.0894$^{+0.0012}_{-0.0009}$    &        0.08573 ($\pm$0.00095)         \\
                HD 190647       &        0.1764$^{+0.0097}_{-0.0082}$   &        0.0891$^{+0.0019}_{-0.0015}$    &        0.0844 ($\pm$0.0015)   &        0.0844 ($\pm$0.0015)    &        0.0801$^{+0.0015}_{-0.0006}$   \\
                HD 190984       &        0.2405$^{+0.0283}_{-0.024}$    &        0.0992$^{+0.0065}_{-0.0054}$    &        0.073$^{+0.0007}_{-0.001}$     &        0.0719$^{+0.0007}_{-0.0009}$    &        0.0718$^{+0.0007}_{-0.0009}$   \\
                HD 204313       &        0.087$^{+0.001}_{-0.042}$      &        0.07238$^{0.00052}_{-0.00061}$  &        0.0696$^{+0.0009}_{-0.0015}$   &        0.0697$^{+0.0008}_{-0.0015}$    &        0.0694$^{+0.0010}_{-0.0015}$   \\
                HD 208487       &        0.2364$^{+0.0087}_{-0.0085}$   &        0.1172 ($\pm$0.0033)    &        0.1013$^{+0.0022}_{-0.0021}$   &        0.1020 ($\pm$0.0022)    &        0.0939 ($\pm$0.0018)   \\
                HD 210702       &        0.531$^{+0.047}_{-0.049}$      &        0.232$^{+0.037}_{-0.033}$       &        0.123$^{+0.015}_{-0.013}$      &        0.113$^{+0.013}_{-0.010}$       &        0.104$^{+0.010}_{-0.008}$      \\
                HD 219077       &        0.299$^{+0.033}_{-0.019}$      &        0.174$^{+0.012}_{-0.010}$       &        X\tnote{1}     &        X\tnote{1}         &        X\tnote{1}     \\
                HD 221287       &        0.615$^{+0.016}_{-0.014}$      &        0.328$^{+0.014}_{-0.015}$       &        0.1417$^{+0.0046}_{-0.0093}$   &        0.1404$^{+0.0046}_{-0.0095}$    &        0.1492$^{+0.0057}_{-0.0083}$   \\
                \hline
        \end{tabular}
        \begin{tablenotes}
\item[1] values are outside image boundaries
\end{tablenotes}
\end{threeparttable}
\end{table*}

\section{Signal-to-noise maps of recovered companion candidates}

\begin{figure*}
\includegraphics[width=\textwidth]{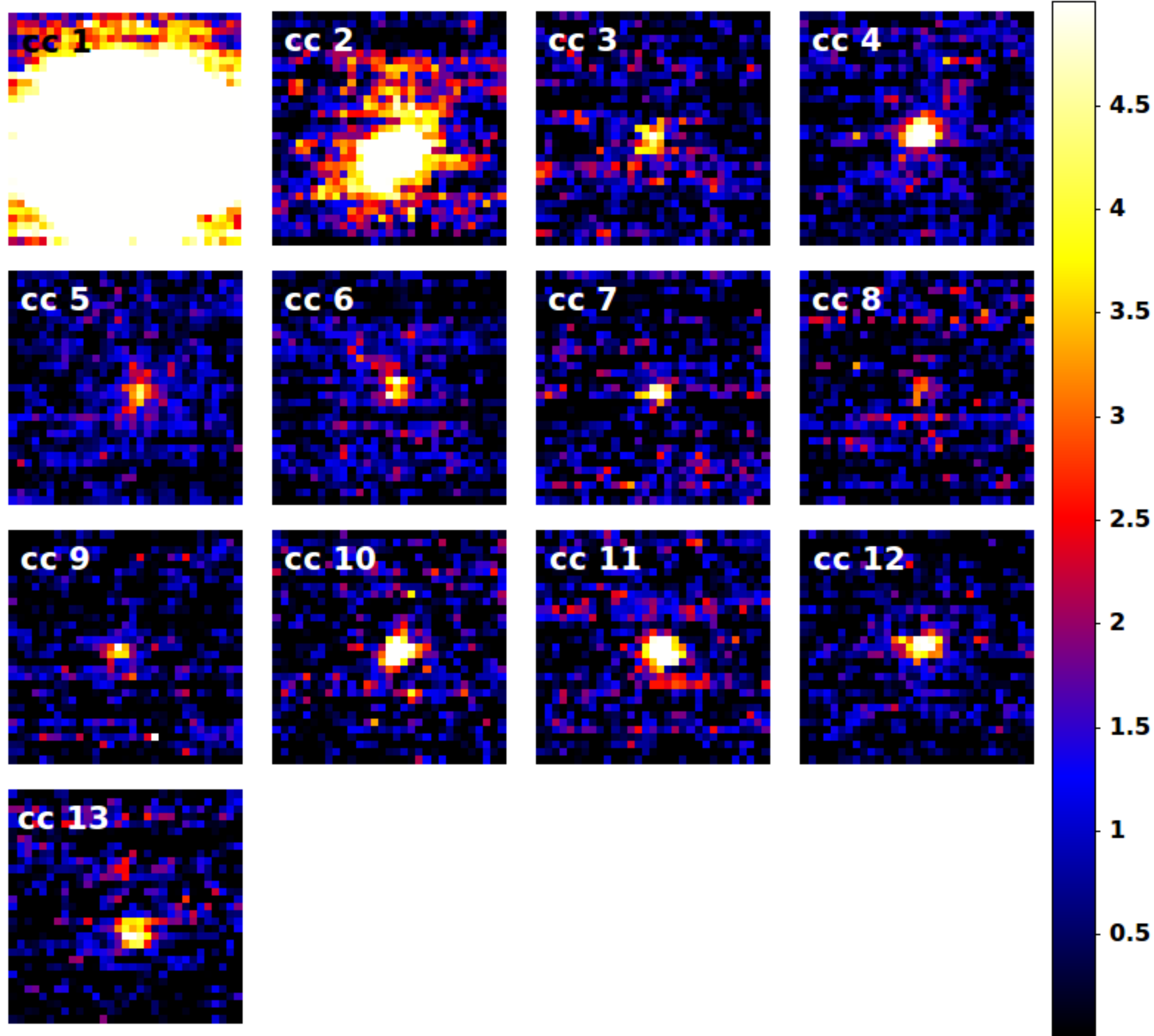}
\caption[]{S/N maps of all detected companion candidates around HD\,85390. The cc numbers are identical to the ones used in Fig.~\ref{fig:companion_candidates_VLT_figure}.} 
\label{sn_maps_hd85390}
\end{figure*}

\begin{figure*}
\includegraphics[width=\textwidth]{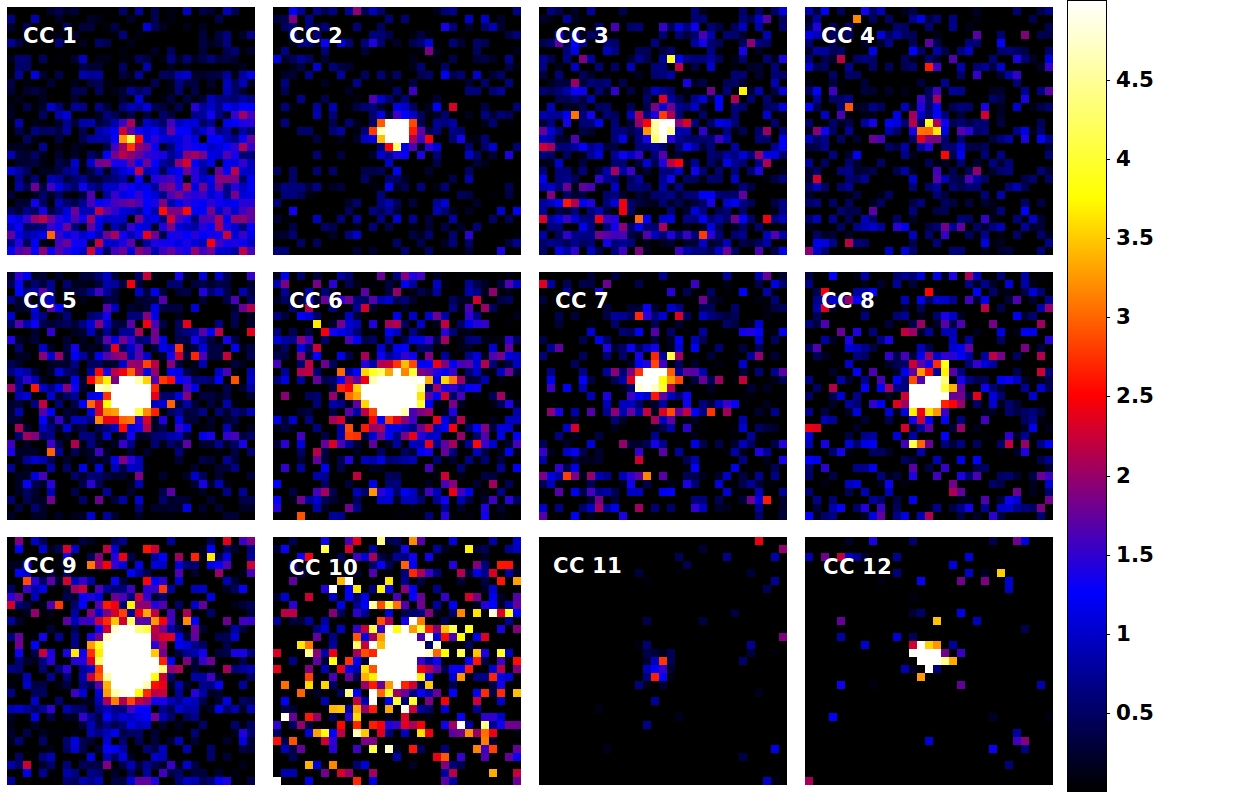}
\caption[]{S/N maps of all detected companion candidates around HD\,113538. The cc numbers are identical to the ones used in Fig.~\ref{fig:companion_candidates_VLT_figure}.} 
\label{sn_maps_hd113538}
\end{figure*}

\clearpage

\section{Contrast curves for all studied targets}

\begin{figure*}
	\captionsetup[subfigure]{labelformat=empty}
        \centering
        \subfloat[GJ 176]{\includegraphics[width=.33\textwidth]{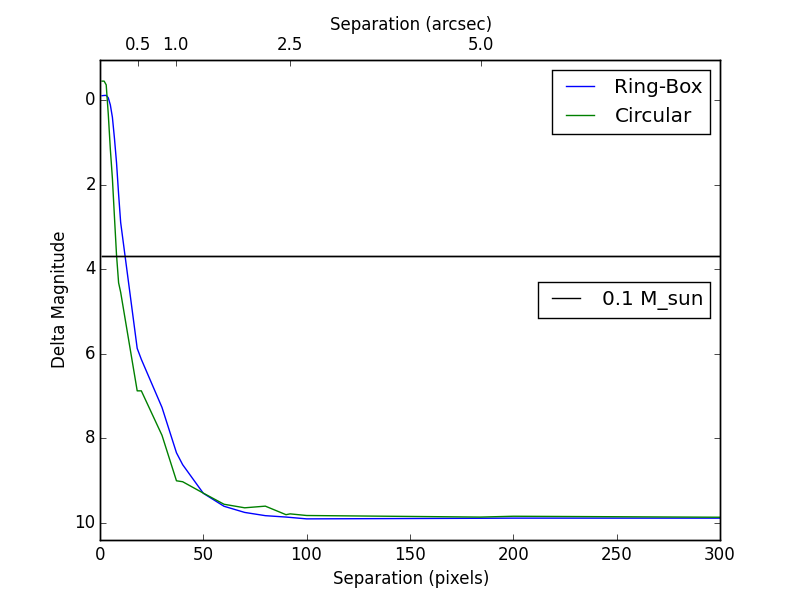}}
        \subfloat[GJ 436]{\includegraphics[width=.33\textwidth]{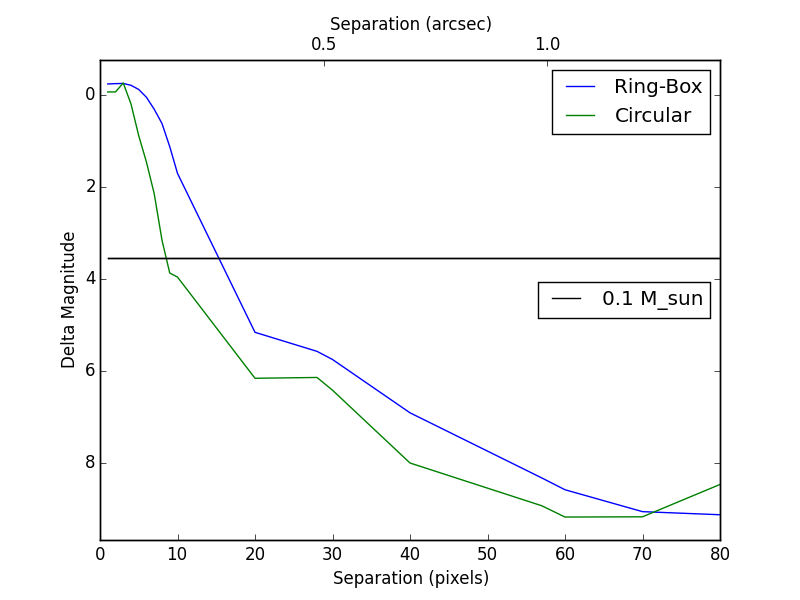}}
        \subfloat[GJ 581]{\includegraphics[width=.33\textwidth]{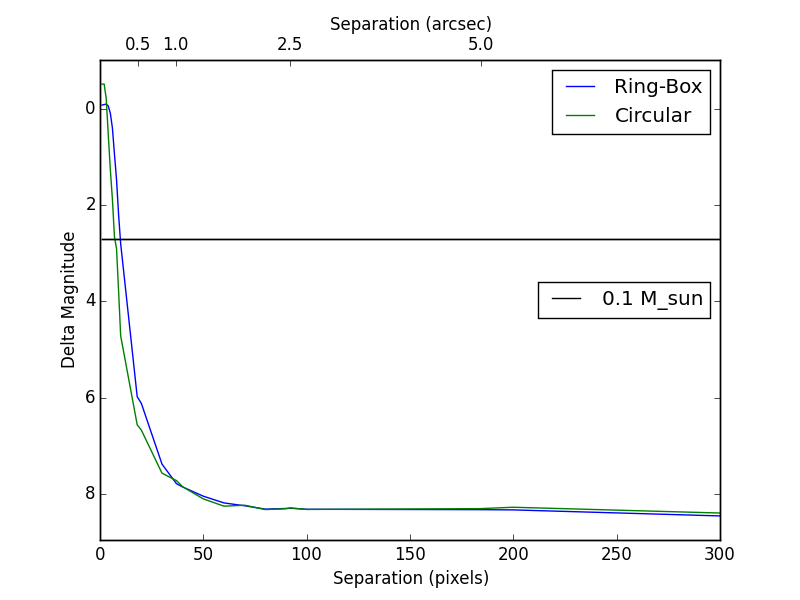}}
        
        \subfloat[GJ 849]{\includegraphics[width=.33\textwidth]{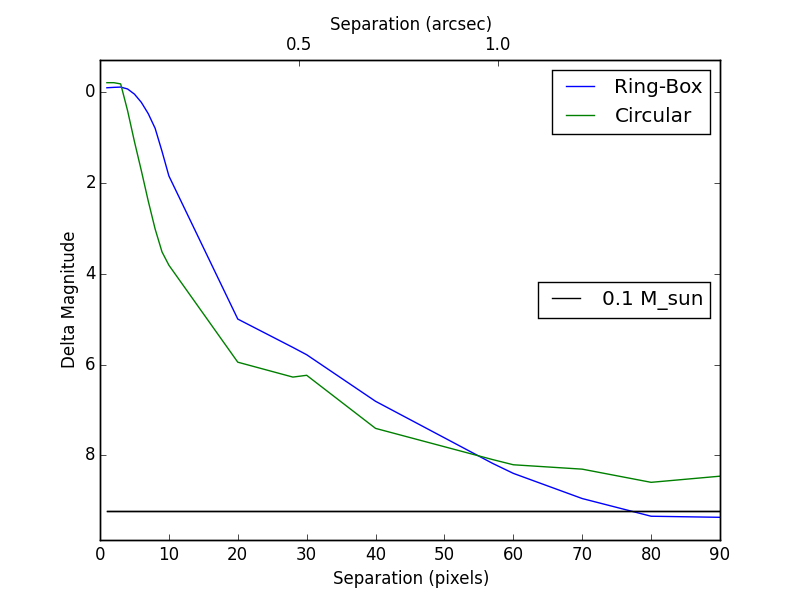}}
        \subfloat[GJ 876]{\includegraphics[width=.33\textwidth]{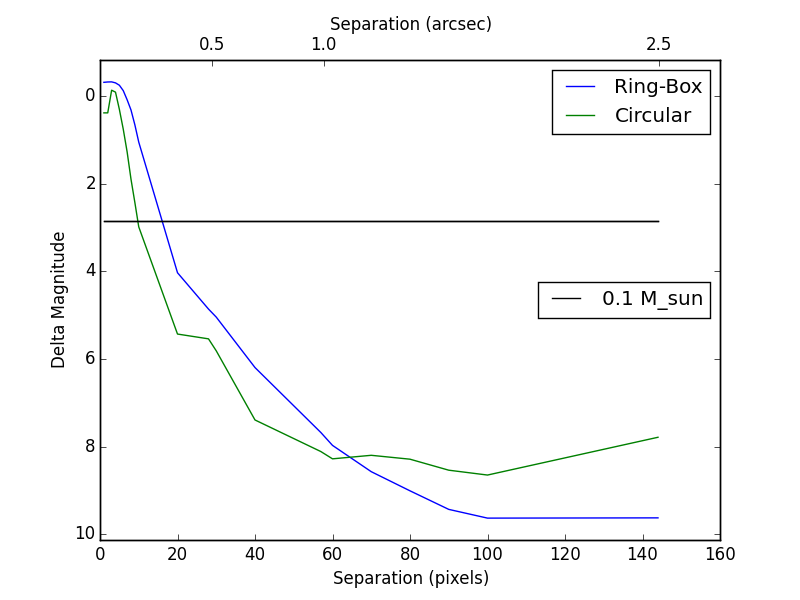}}
        \subfloat[HD 4113]{\includegraphics[width=.33\textwidth]{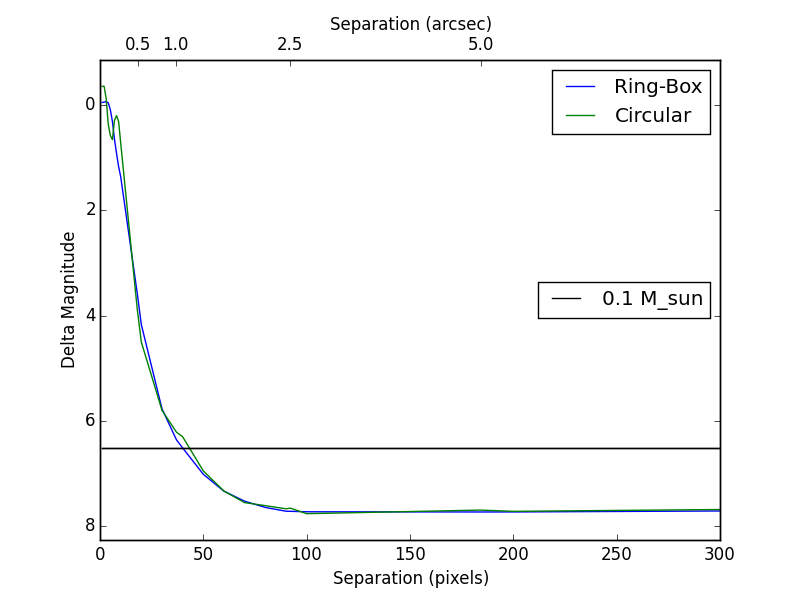}}
        
        \subfloat[HD 10647]{\includegraphics[width=.33\textwidth]{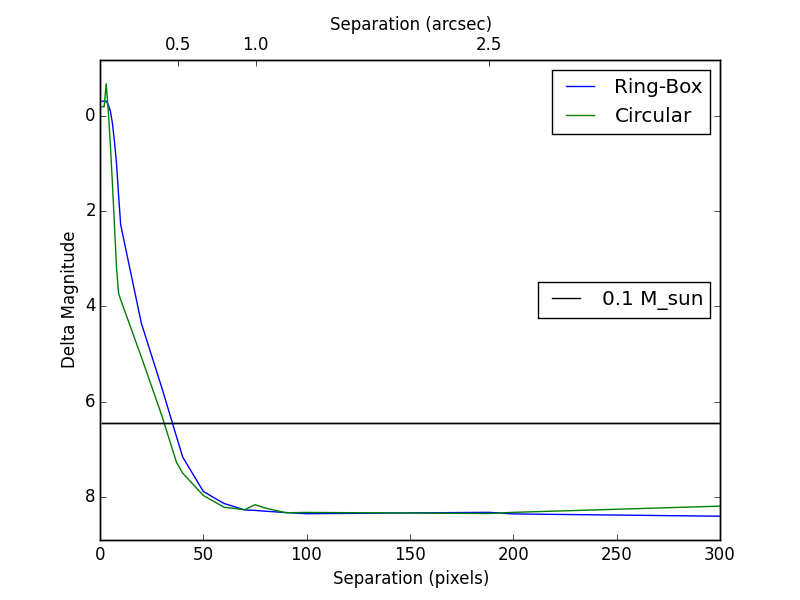}}
        \subfloat[HD 11506]{\includegraphics[width=.33\textwidth]{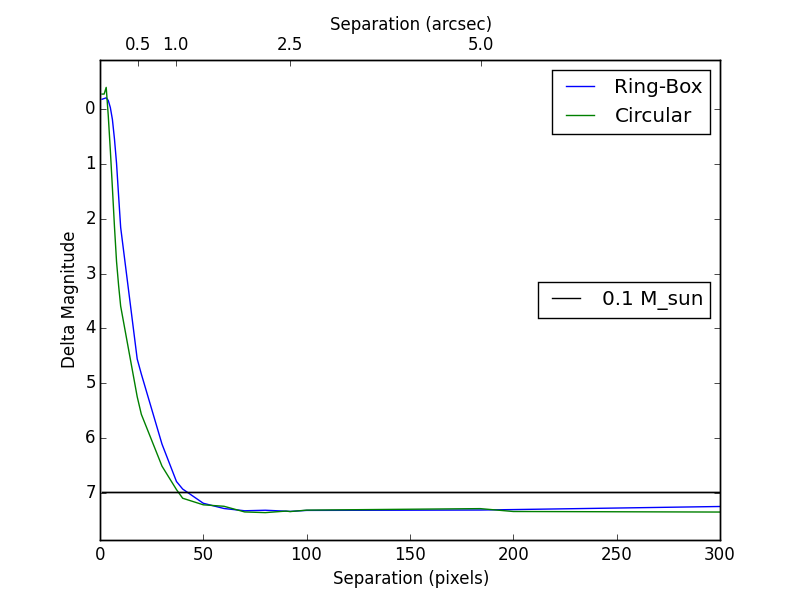}}
        \subfloat[HD 11964]{\includegraphics[width=.33\textwidth]{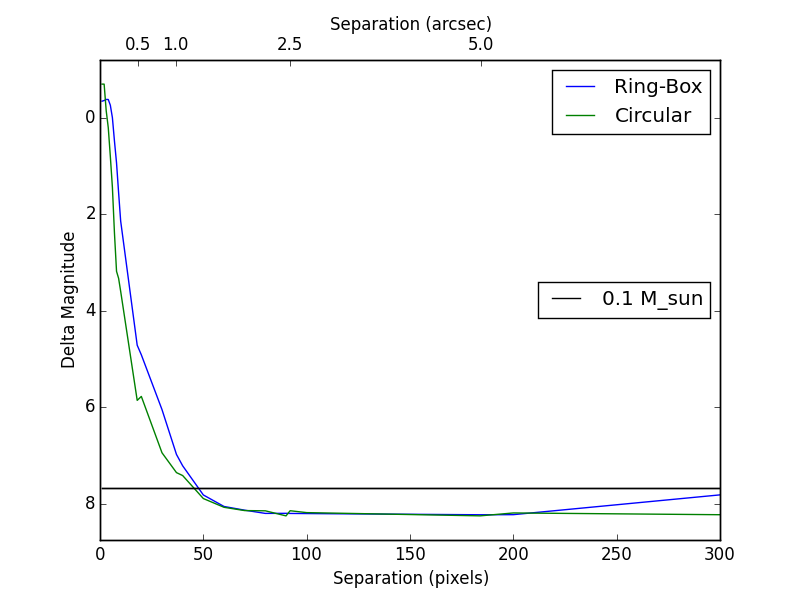}}
        
        \subfloat[HD 20782]{\includegraphics[width=.33\textwidth]{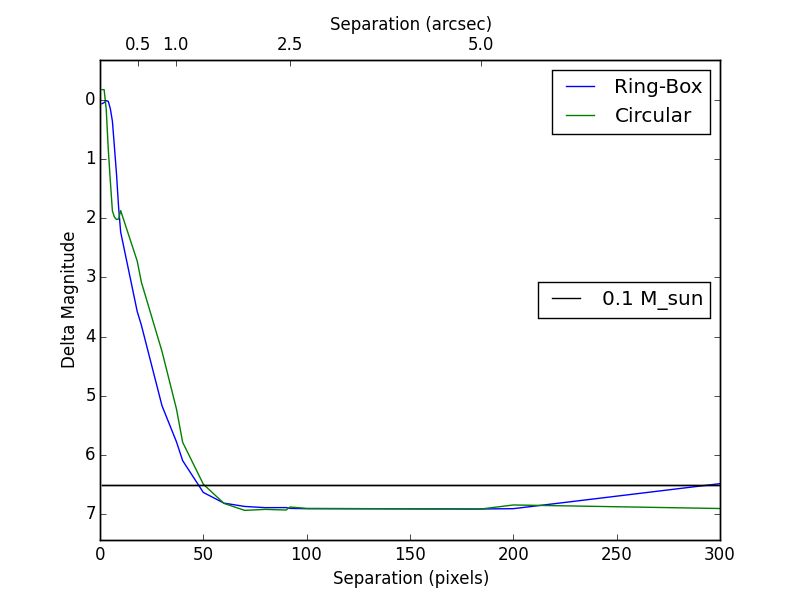}}
        \subfloat[HD 21693]{\includegraphics[width=.33\textwidth]{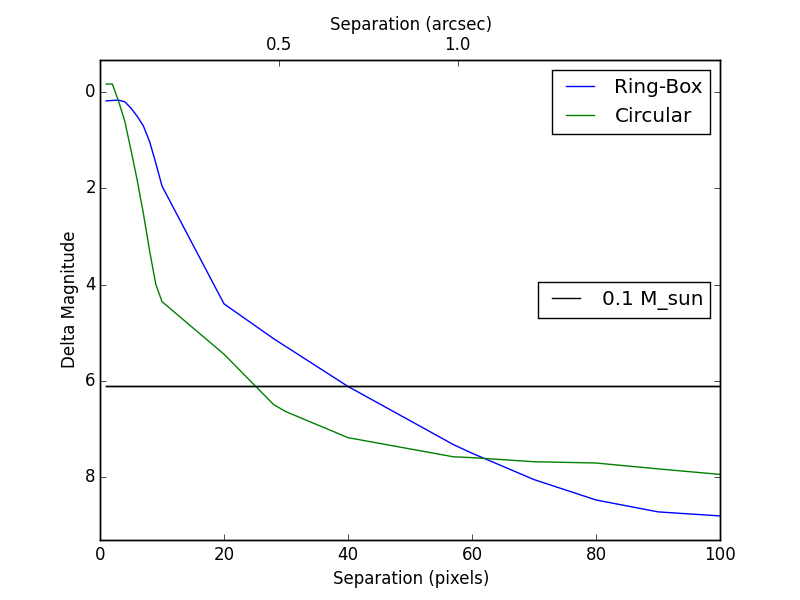}}
        \subfloat[HD 23127]{\includegraphics[width=.33\textwidth]{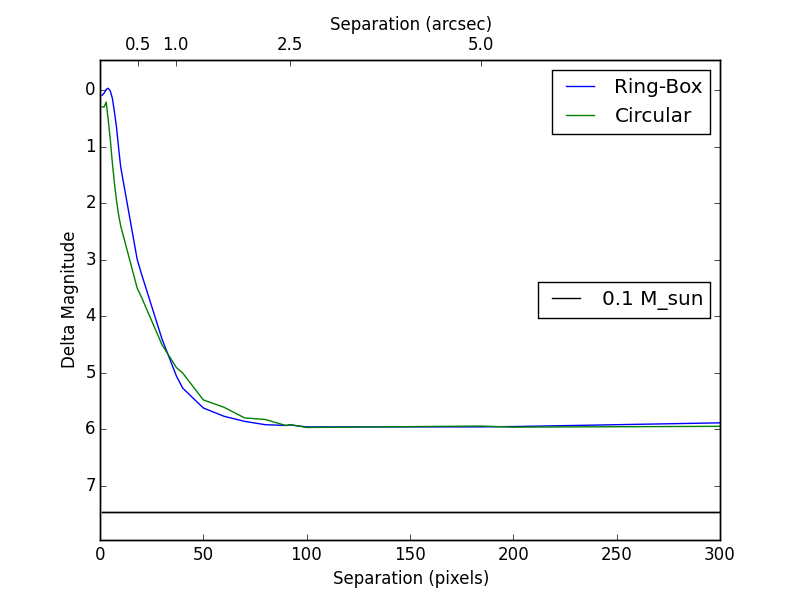}}
        
        \phantomcaption
        \phantomcaption
        \end{figure*}
        \begin{figure*}
        \ContinuedFloat
        \captionsetup[subfigure]{labelformat=empty}
                
        \subfloat[HD 30562]{\includegraphics[width=.33\textwidth]{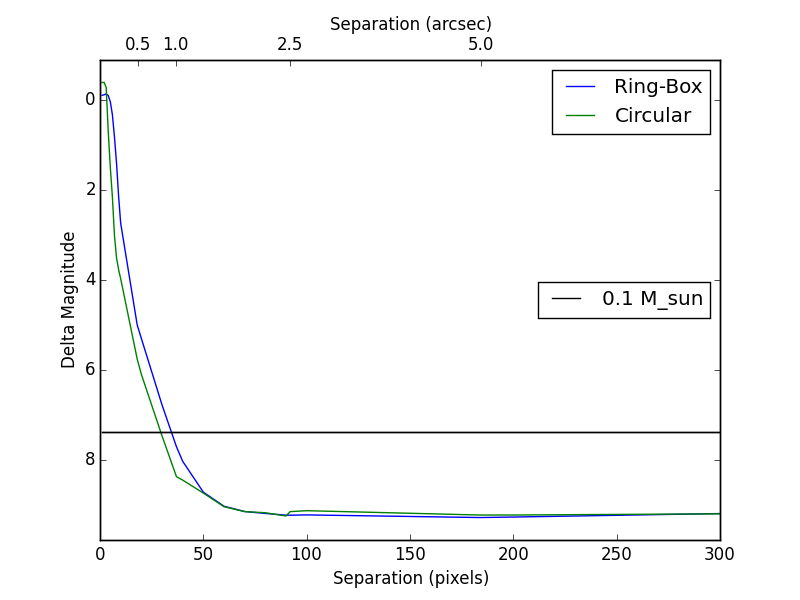}}
        \subfloat[HD 33283]{\includegraphics[width=.33\textwidth]{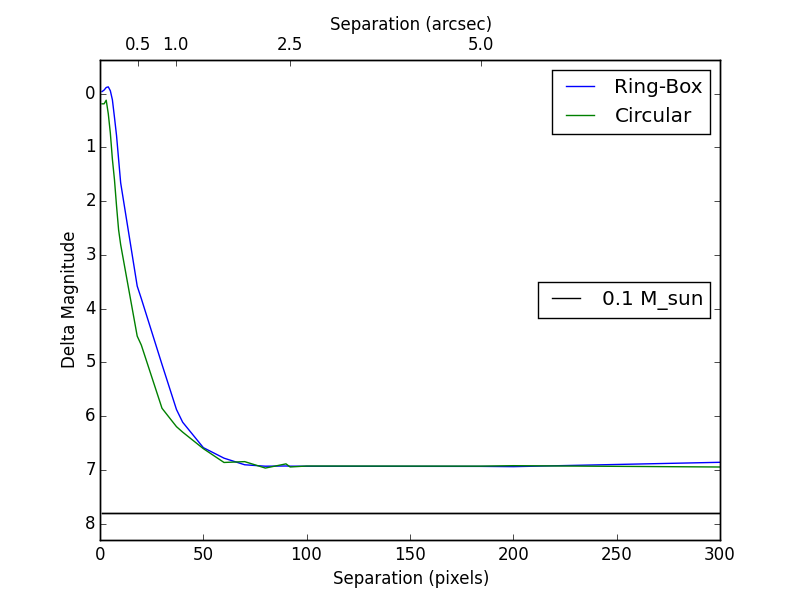}}
        \subfloat[HD 47186]{\includegraphics[width=.33\textwidth]{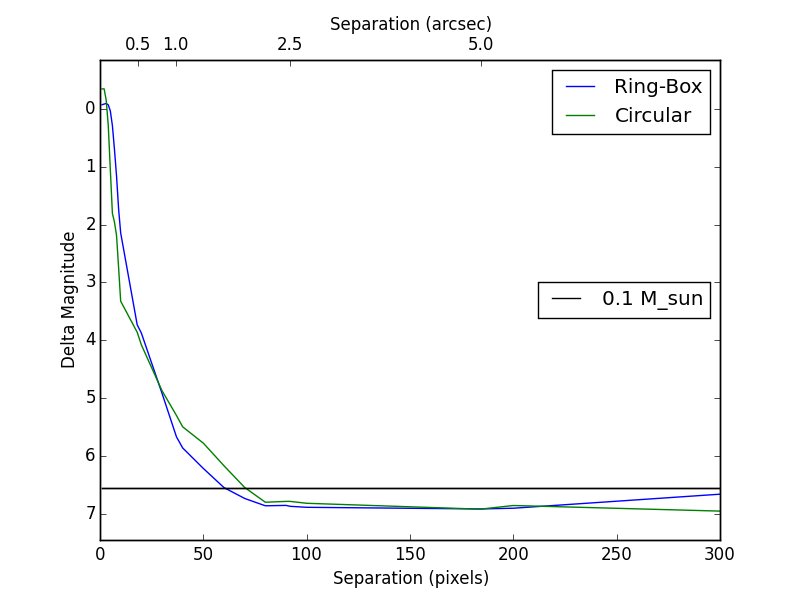}}       
        
        \subfloat[HD 60532]{\includegraphics[width=.33\textwidth]{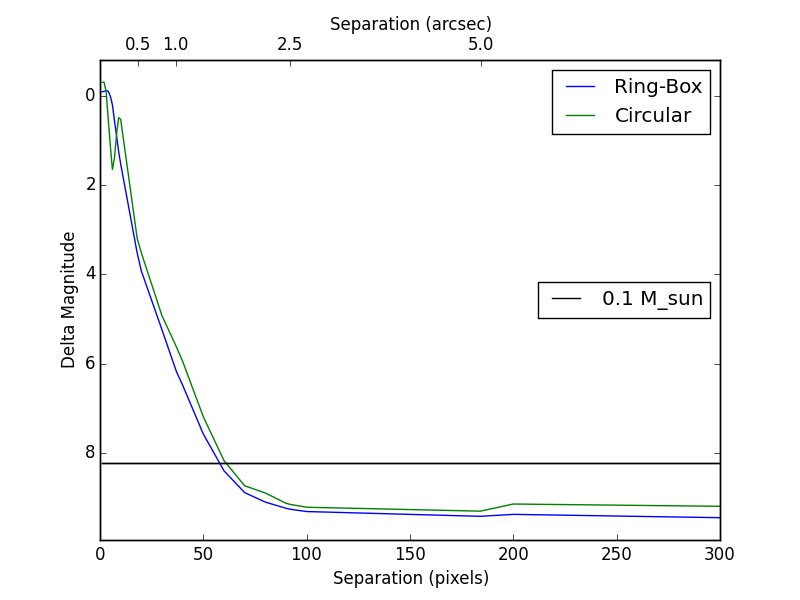}}
        \subfloat[HD 69830]{\includegraphics[width=.33\textwidth]{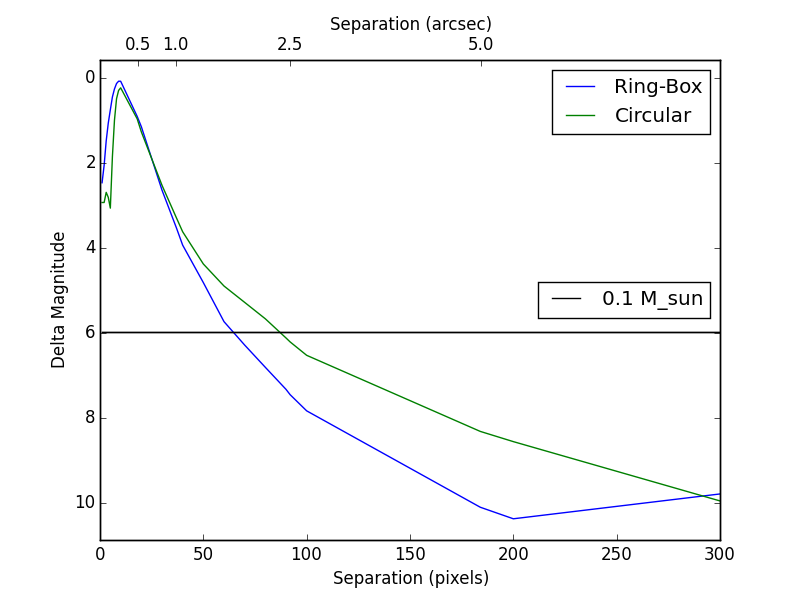}}
        \subfloat[HD 85390]{\includegraphics[width=.33\textwidth]{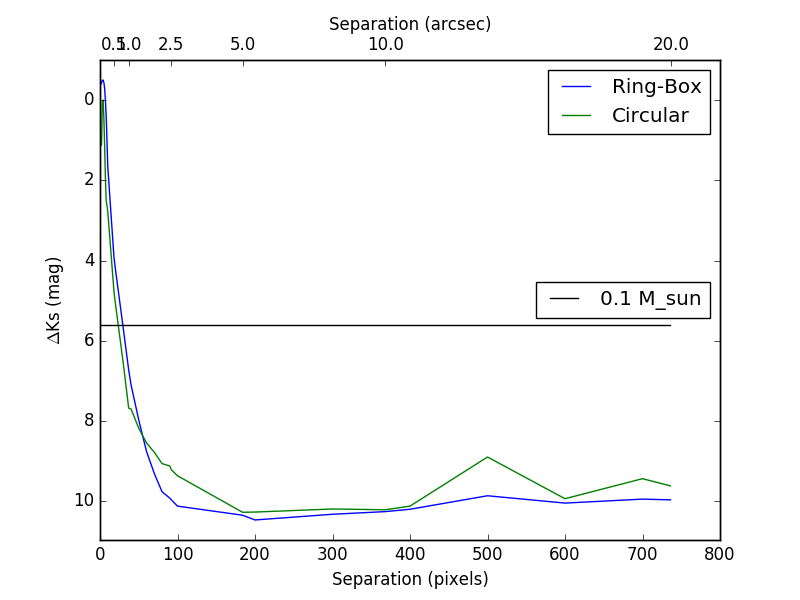}}
        
        \subfloat[HD 85512]{\includegraphics[width=.33\textwidth]{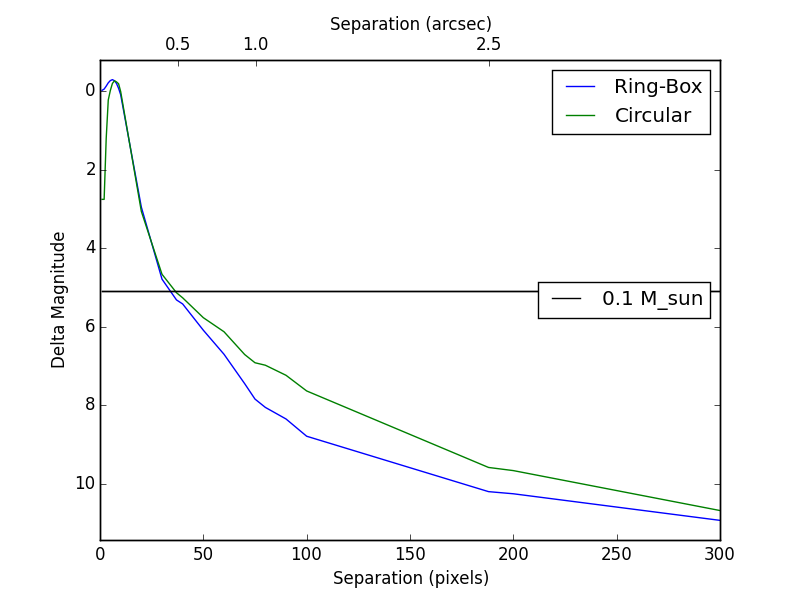}}
        \subfloat[HD 98649]{\includegraphics[width=.33\textwidth]{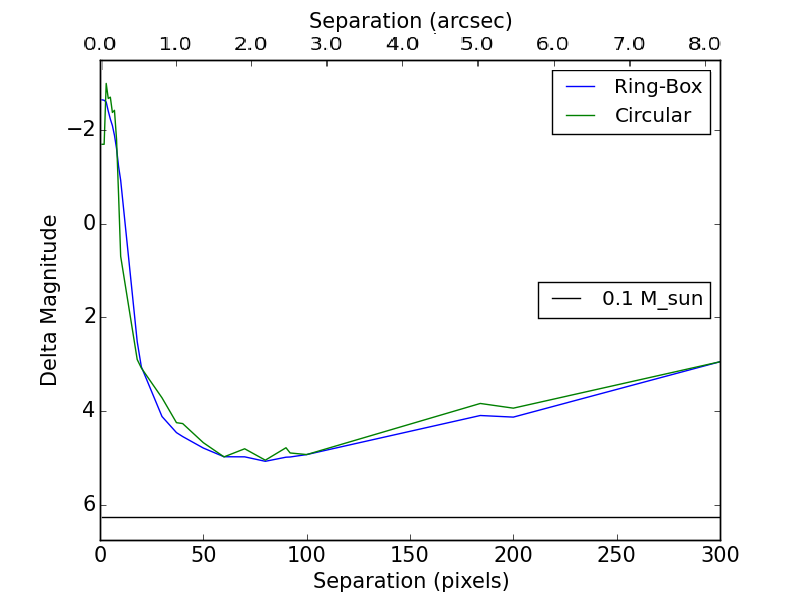}}
        \subfloat[HD 113538]{\includegraphics[width=.33\textwidth]{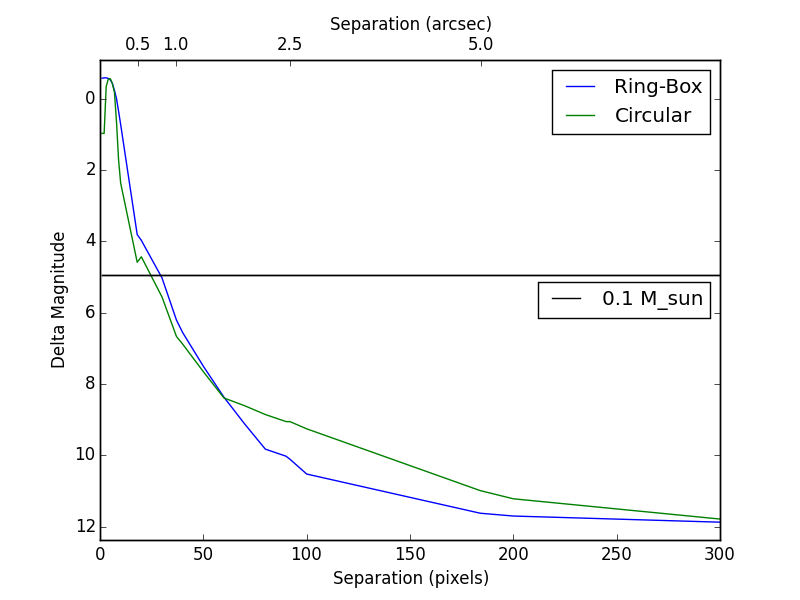}}
        
        \subfloat[HD 114613]{\includegraphics[width=.33\textwidth]{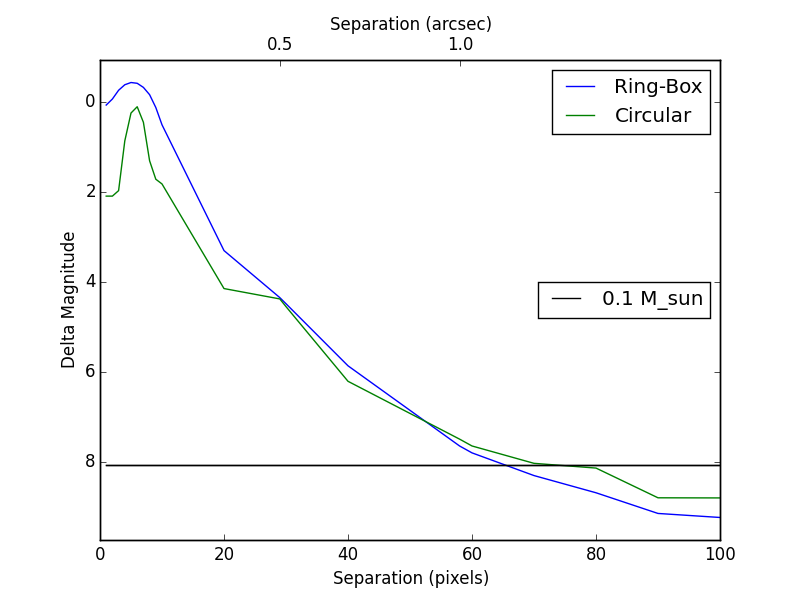}}
        \subfloat[HD 142022]{\includegraphics[width=.33\textwidth]{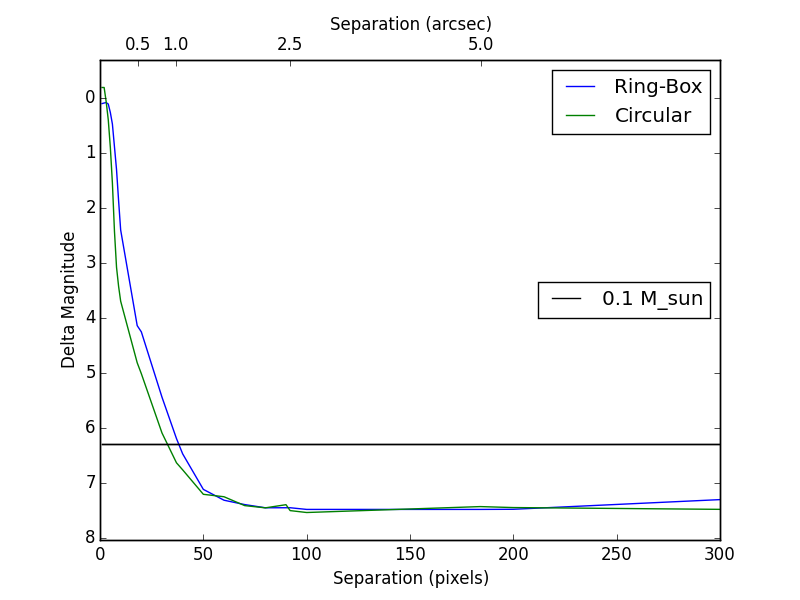}}
        \subfloat[HD 147873]{\includegraphics[width=.33\textwidth]{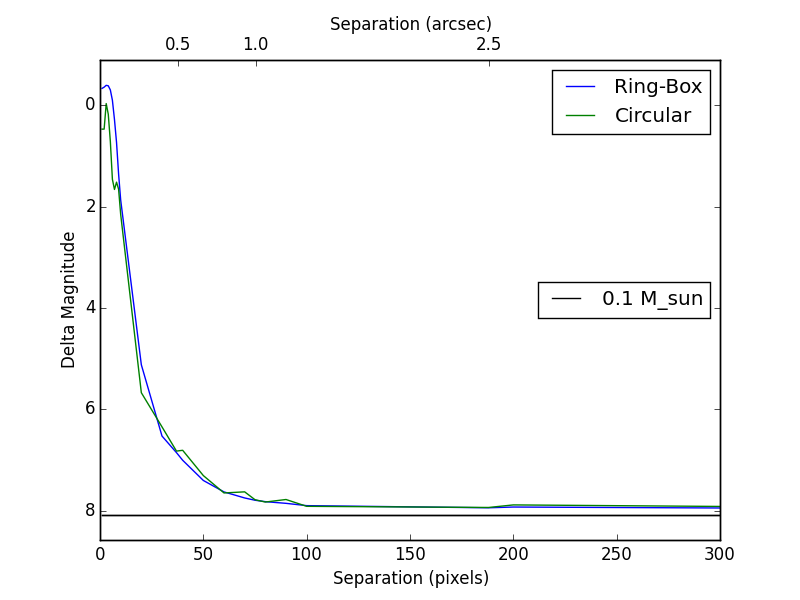}}
        
        \phantomcaption
        \phantomcaption
        \end{figure*}
        \begin{figure*}
        \ContinuedFloat
	\captionsetup[subfigure]{labelformat=empty}
                
        \subfloat[HD 154857]{\includegraphics[width=.33\textwidth]{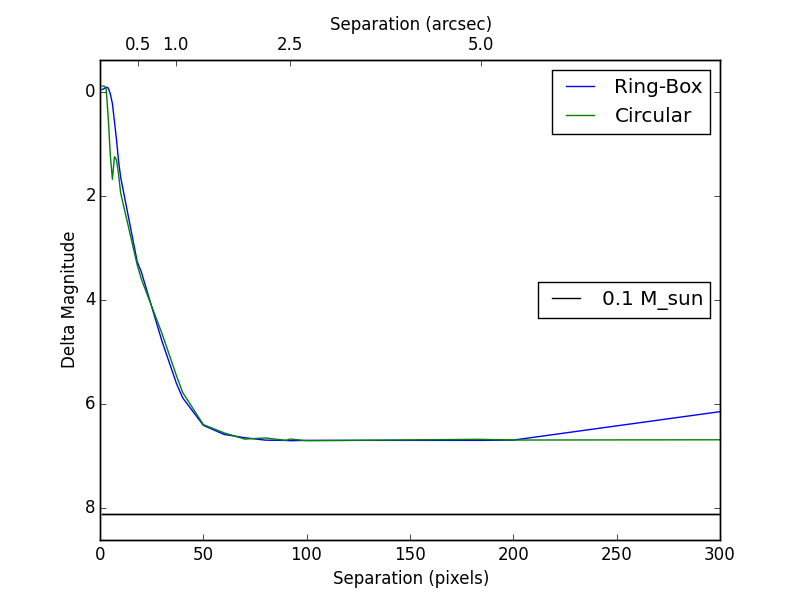}}     
        \subfloat[HD 159868]{\includegraphics[width=.33\textwidth]{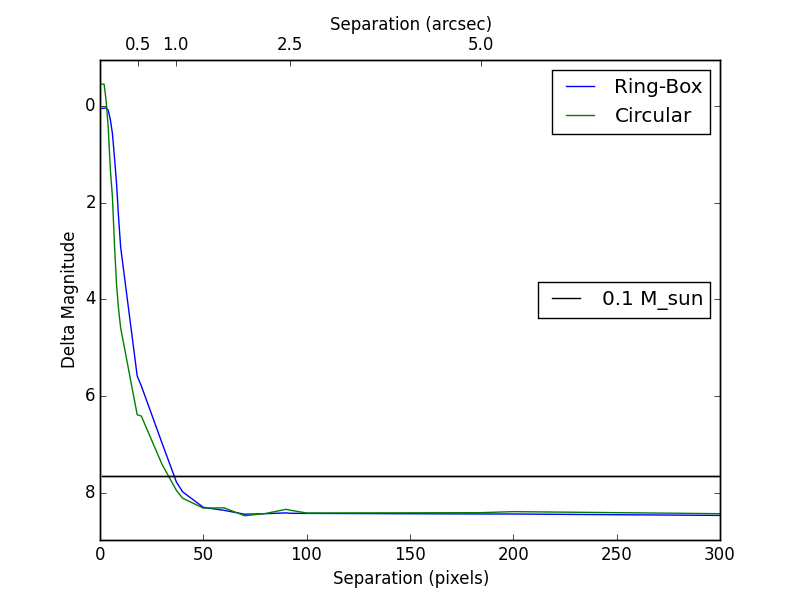}}
        \subfloat[HD 168443]{\includegraphics[width=.33\textwidth]{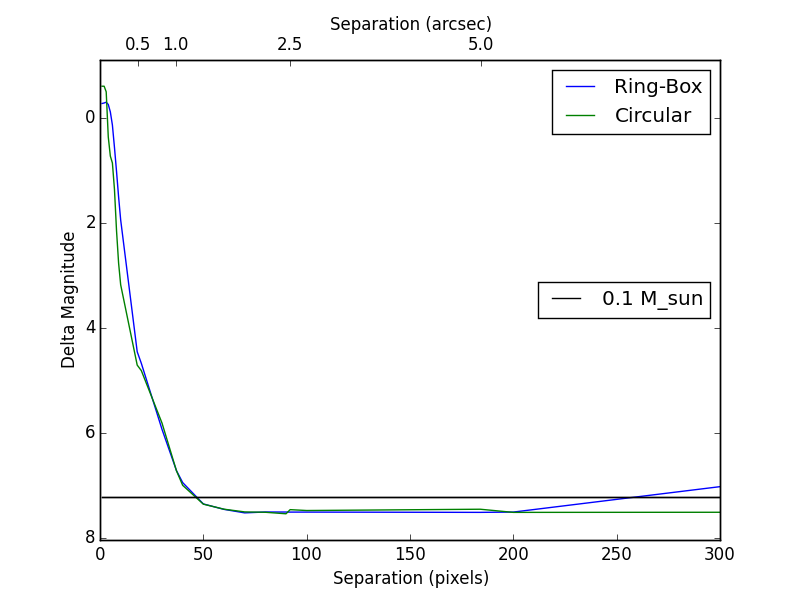}}
        
        \subfloat[HD 171238]{\includegraphics[width=.33\textwidth]{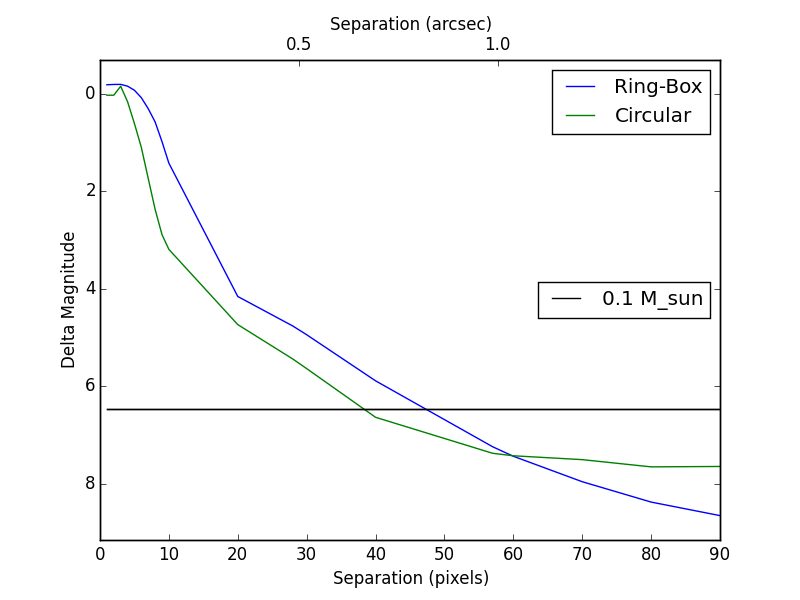}}
        \subfloat[HD 175167]{\includegraphics[width=.33\textwidth]{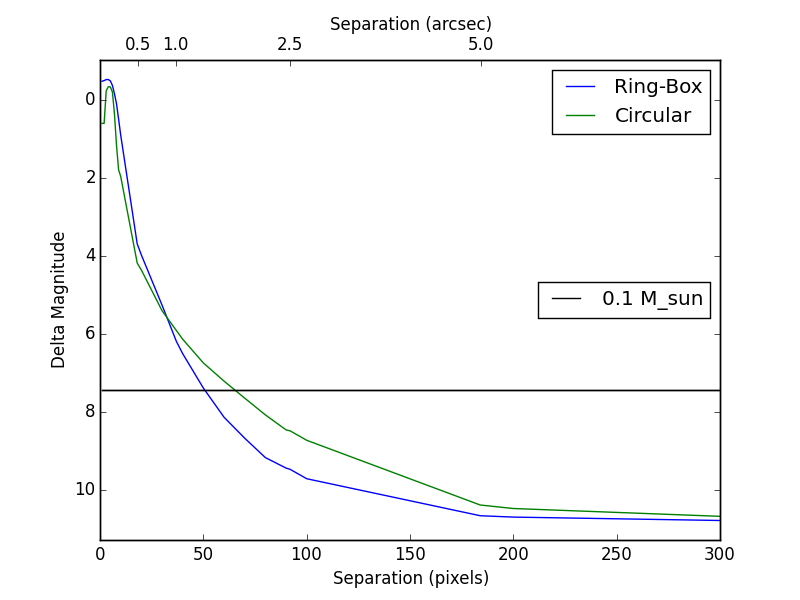}}
        \subfloat[HD 183263]{\includegraphics[width=.33\textwidth]{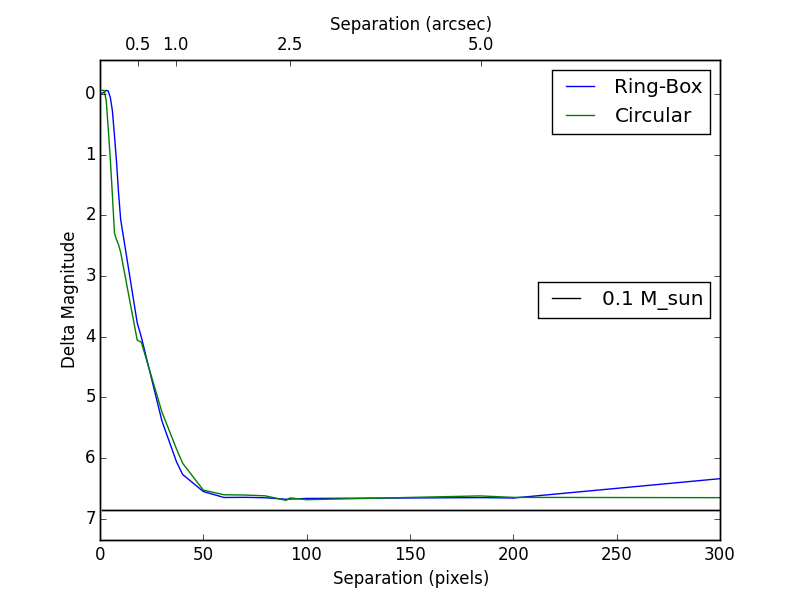}}             
        
        \subfloat[HD 187085]{\includegraphics[width=.33\textwidth]{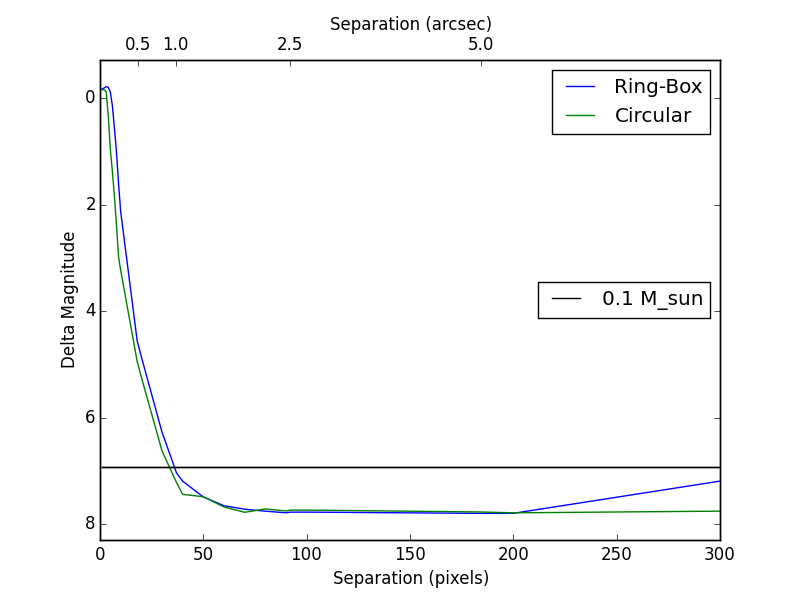}}     
        \subfloat[HD 190647]{\includegraphics[width=.33\textwidth]{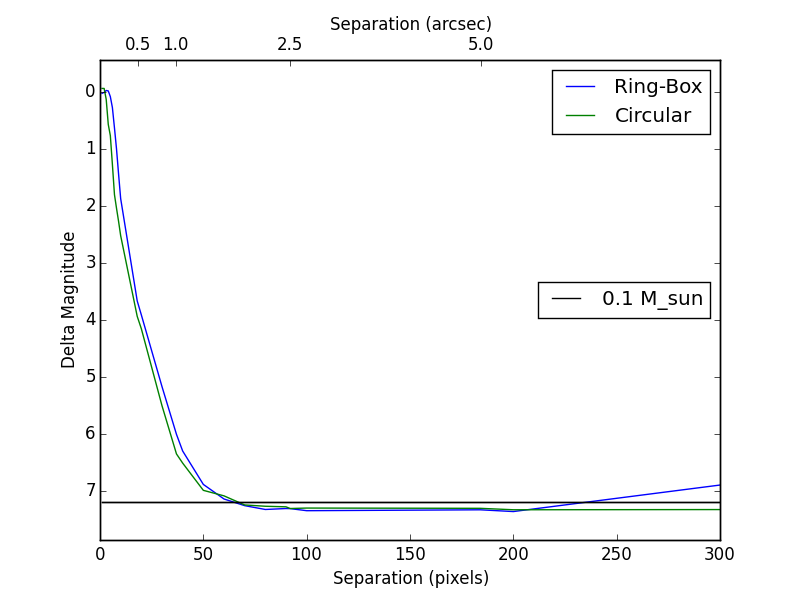}}
        \subfloat[HD 190984]{\includegraphics[width=.33\textwidth]{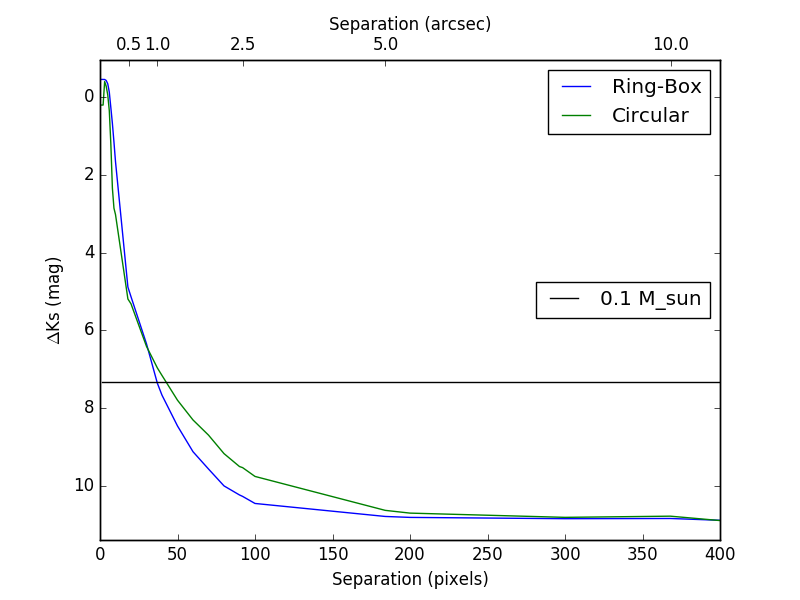}}
        
        \subfloat[HD 204313]{\includegraphics[width=.33\textwidth]{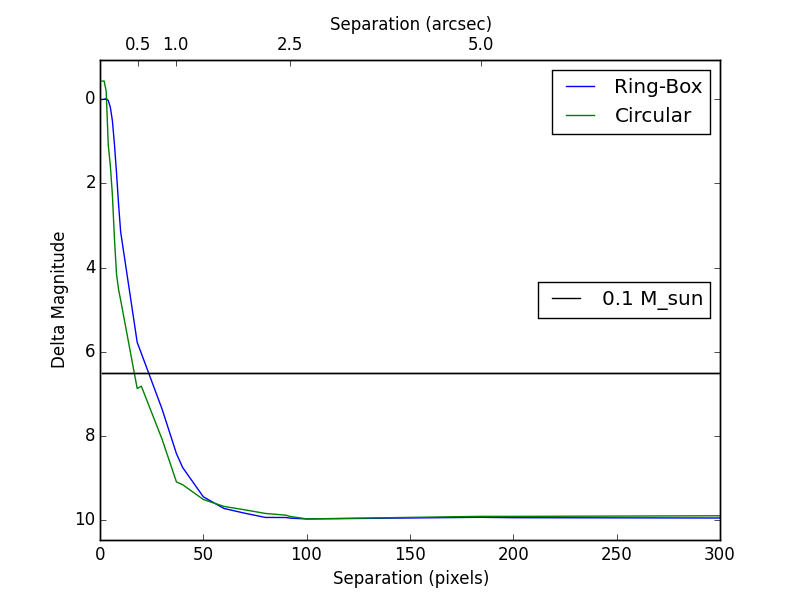}}
        \subfloat[HD 208487]{\includegraphics[width=.33\textwidth]{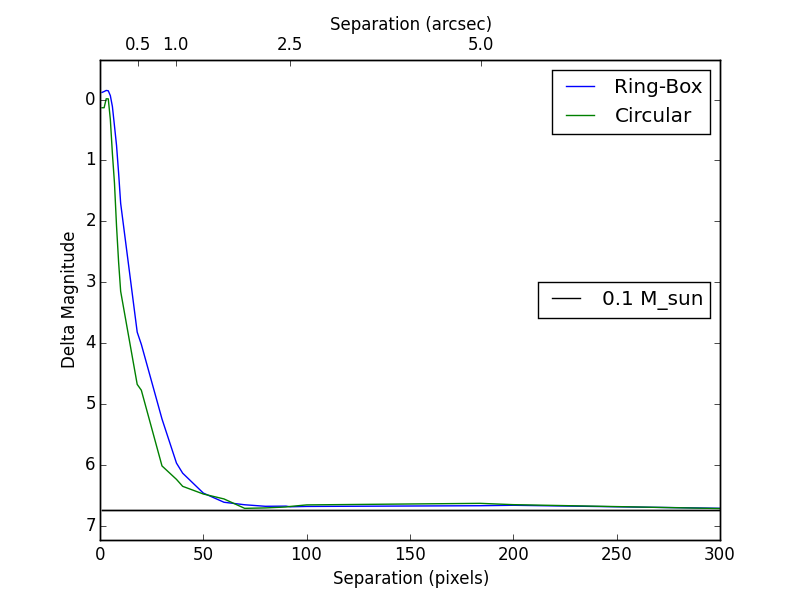}}
        \subfloat[HD 210702]{\includegraphics[width=.33\textwidth]{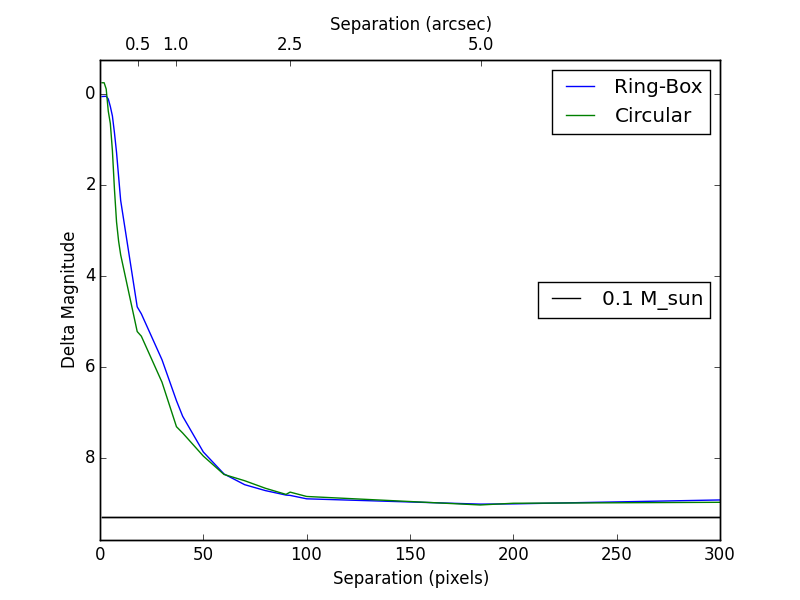}}
        
        \phantomcaption
        \phantomcaption
        \end{figure*}
        \begin{figure*}
        \ContinuedFloat
        \captionsetup[subfigure]{labelformat=empty}
        
        \subfloat[HD 219077]{\includegraphics[width=.33\textwidth]{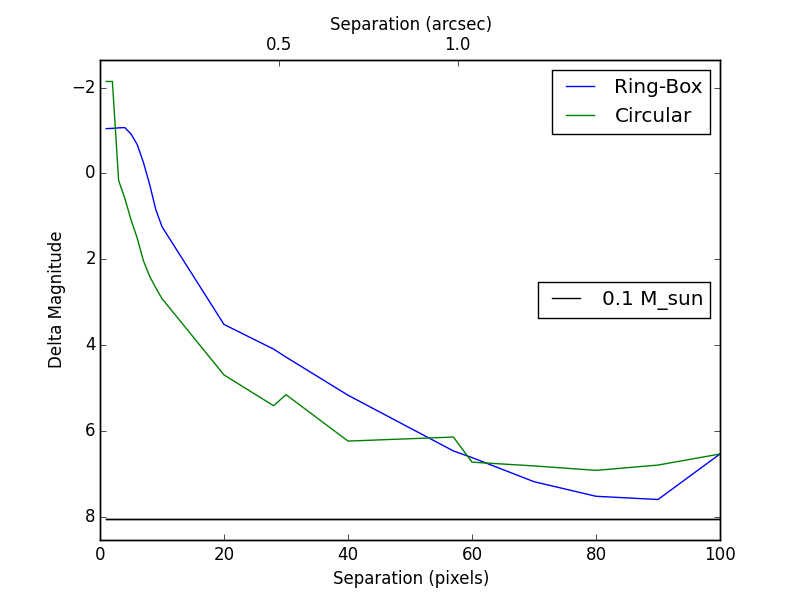}}
        \subfloat[labelformat=empty][HD 221287]{\includegraphics[width=.33\textwidth]{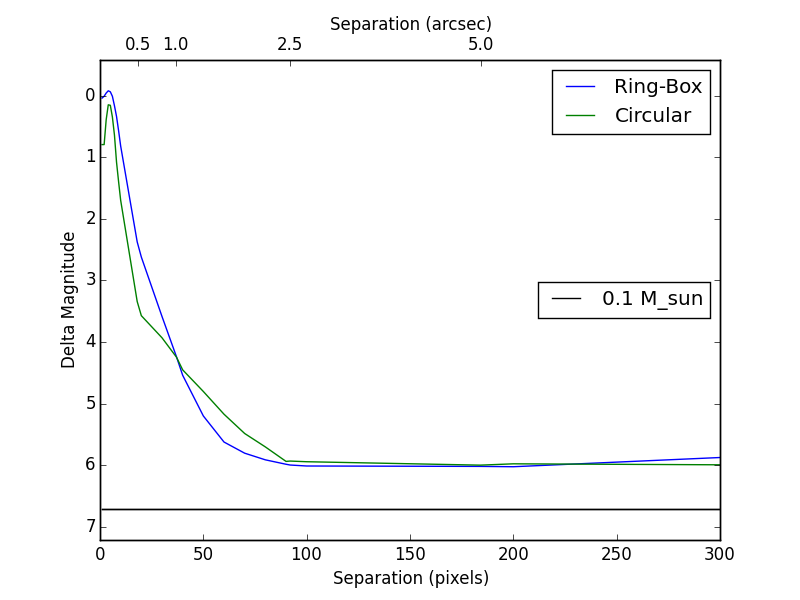}}
        
    \caption{Magnitude detection limits for each of the systems we studied.}
    \label{fig:detection_limits_figure}
\end{figure*}

\begin{figure}
\includegraphics[scale=0.42]{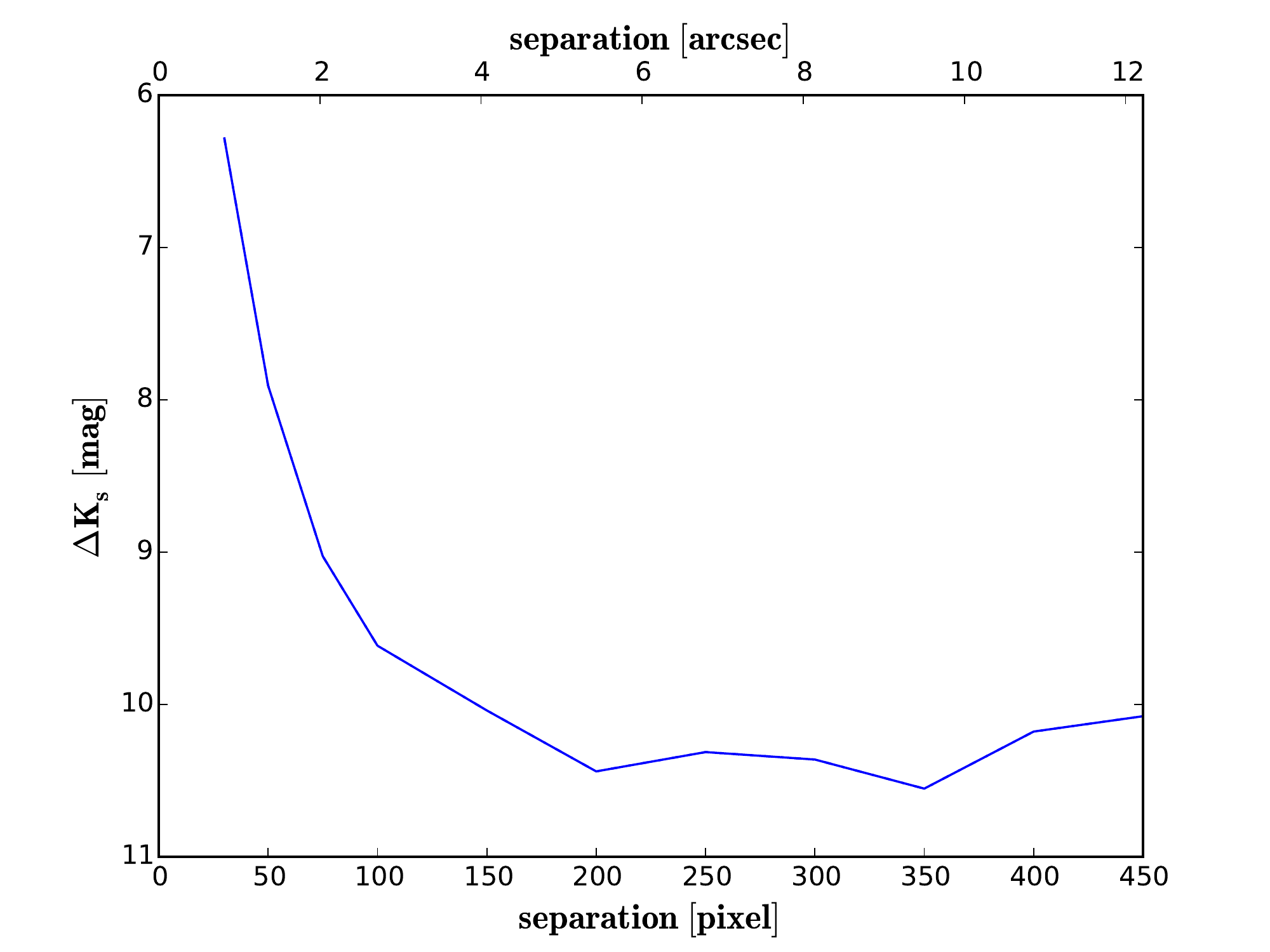}
\caption[]{ADI 5$\sigma$ detection limits of HD\,40307. The detection limits were computed by inserting fake companions in the dataset, to account for the ADI self subtraction effect.} 
\label{adi-contrast}
\end{figure}

\end{appendix}
\end{document}